\documentclass[%
reprint,
superscriptaddress,
amsmath,amssymb,
aps,
pra,
]{revtex4-2}

\usepackage{graphicx}
\usepackage{dcolumn}
\usepackage{bm}
\usepackage{bbm}
\usepackage{braket}
\usepackage{comment}
\usepackage[dvipsnames]{xcolor}
\usepackage{hyperref}
\usepackage[resetlabels,labeled]{multibib}
\newcites{First}{References}

\definecolor{darkblue}{rgb}{0.,0.,0.6}
\definecolor{darkgreen}{rgb}{0.,0.6,0.0}
\definecolor{darkorange}{rgb}{0.8,0.4,0.2}

\newcommand{\Yb}{$^{171}$Yb$^+$\,}
\newcommand{\Ybc}{$^{172}$Yb$^+$\,}

\begin{document}
\preprint{APS/123-QED}

\title{Trapped-Ion Quantum Simulation of Electron \\ Transfer Models with Tunable Dissipation}
\author{Visal So}
\thanks{These authors contributed equally to this work.}
\affiliation{Department of Physics and Astronomy, Rice University, Houston, TX 77005, USA}
\author{Midhuna Duraisamy Suganthi}
\thanks{These authors contributed equally to this work.}
\affiliation{Department of Physics and Astronomy, Rice University, Houston, TX 77005, USA}
\affiliation{Applied Physics Graduate Program, Smalley-Curl Institute, Rice University, Houston, TX 77005, USA }
\author{Abhishek Menon}
\affiliation{Department of Physics and Astronomy, Rice University, Houston, TX 77005, USA}
\author{Mingjian Zhu}
\affiliation{Department of Physics and Astronomy, Rice University, Houston, TX 77005, USA}
\author{Roman Zhuravel}
\affiliation{Department of Physics and Astronomy, Rice University, Houston, TX 77005, USA}
\author{Han Pu}
\affiliation{Department of Physics and Astronomy, Rice University, Houston, TX 77005, USA}
\author{Peter G. Wolynes}
\affiliation{Department of Physics and Astronomy, Rice University, Houston, TX 77005, USA}
\affiliation{Department of Chemistry, Rice University, Houston, TX 77005, USA}
\affiliation{Center for Theoretical Biological Physics, Rice University, Houston, TX 77005, USA}
\affiliation{Department of Biosciences, Rice University, Houston, TX 77005, USA}
\author{José N. Onuchic}
\affiliation{Department of Physics and Astronomy, Rice University, Houston, TX 77005, USA}
\affiliation{Department of Chemistry, Rice University, Houston, TX 77005, USA}
\affiliation{Center for Theoretical Biological Physics, Rice University, Houston, TX 77005, USA}
\affiliation{Department of Biosciences, Rice University, Houston, TX 77005, USA}
\author{Guido Pagano}
\email{pagano@rice.edu}
\affiliation{Department of Physics and Astronomy, Rice University, Houston, TX 77005, USA}


\begin{abstract}
Electron transfer is at the heart of many fundamental physical, chemical, and biochemical processes essential for life. The exact simulation of these reactions is often hindered by the large number of degrees of freedom and by the essential role of quantum effects. Here, we experimentally simulate a paradigmatic model of molecular electron transfer using a multispecies trapped-ion crystal, where the donor-acceptor gap, the electronic and vibronic couplings, and the bath relaxation dynamics can all be controlled independently. By manipulating both the ground-state and optical qubits, we observe the real-time dynamics of the spin excitation, measuring the transfer rate in several regimes of adiabaticity and relaxation dynamics. Our results provide a testing ground for increasingly rich models of molecular excitation transfer processes that are relevant for molecular electronics and light-harvesting systems.
\end{abstract}
\maketitle

\section*{Introduction}

Quantum devices hold the promise to provide an advantage in directly simulating many-body quantum systems \cite{Daley2022}. Chemical reaction dynamics provides a wide range of target applications. Fully realistic digitization of the real-time dynamics of molecules on fault-tolerant quantum computers, however, requires qubit numbers and circuit depths that exceed the current state of the art \cite{Reiher2017}. A promising alternative approach is to develop programmable analog quantum simulators \cite{maskara2023programmable, Mostame2016, Kang2024} that map the dynamical degrees of freedom of molecules directly onto the quantum hardware, therefore providing a more direct but problem-specific quantum advantage.

One outstanding challenge is modeling the real-time electron transfer (ET) dynamics in molecular systems embedded in biological environments. In these systems, the energy differences between the electronic states, molecular vibrational energies, and their mutual couplings are all of the same order of magnitude. This requires simulating electronic excitations while taking into account a large number of nuclear degrees of freedom. In addition, reactions at low temperatures in many molecular systems, ranging from myoglobin ligand recombination \cite{Frauenfelder1985} to charge transport in DNA strands \cite{Zhuravel2020}, suggest that quantum effects play a key role.

In many regimes, the reaction dynamics can be treated using imaginary-time path-integral methods {\cite{Wolynes1987a, Zheng1989, Lawrence2020}. It has also proven expedient to treat the nuclear and electronic degrees of freedom using a mix of quantum and classical dynamics \cite{Bittner1995}, but the limits of this approach are not always clear. When quantum coherences between the electronic and vibrational degrees of freedom \cite{Scholes2017, Wang2019} are relevant, such approaches are only approximate. Methods based on the hierarchical equations of motion approach \cite{Tanimura2020}, tensor networks \cite{Strathearn2018, Tamascelli2019, Somoza2019}, and real-time path-integral evaluations \cite{Makri2018, Kundu2020} have also made progress in those regimes.

Recently, the high degree of control and tunability of programmable quantum platforms such as trapped ions, superconducting qubits, and photonic simulators have been used to experimentally simulate models of vibrationally assisted energy transfer \cite{Gorman2018}, conical intersections \cite{Whitlow2023, Valahu2023, Wang2023}, noise-assisted excitation transfer \cite{Potonik2018, Maier2019}, ET driven by polarized light \cite{Ke2023}, and molecular vibrational dynamics \cite{Sparrow2018}.

In this work, we show that a trapped-ion quantum simulator with independent control of unitary and dissipative processes can successfully simulate a paradigmatic ET model. This is achieved by manipulating two different atomic ion species and using both ground-state and optical qubits, combining spin and spin-motion coherent manipulation with sympathetic cooling \cite{Rohde2001cool, Blinov2002cool} of a collective bosonic mode. This programmable open quantum system enables the measurement of the time-resolved dynamics of the system in contact with an engineered bosonic bath, accessing nonperturbative regimes, where electronic and vibrational excitations, their mutual coupling, and the relaxation rate are all of the same order of magnitude.

\section*{Results}

\begin{figure*}[t!]
    \centering
    \includegraphics[width=\linewidth]{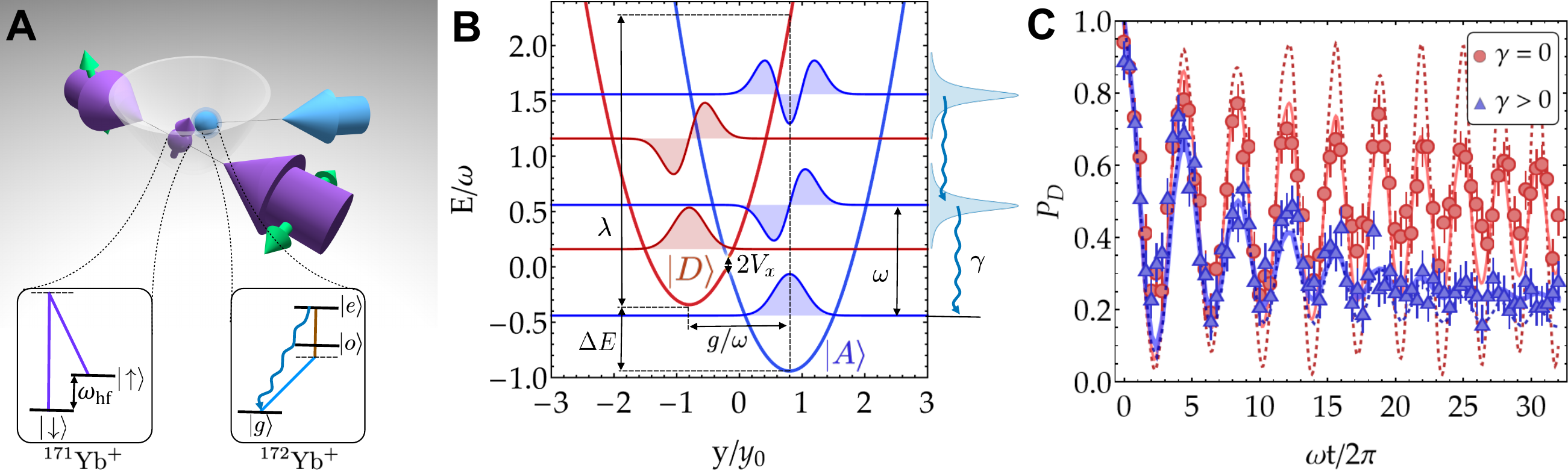}
    \caption{
    {\bf Simulating ET model with tunable dissipation.} \\({\bf A}) \Yb-\Ybc ion crystal confined in a harmonic potential with Coulomb interactions defining normal modes of motion. The ground-state qubit of \Yb encodes the spin degree of freedom and is coherently manipulated by two counterpropagating 355-nm Raman beams (purple arrows, with green arrows showing the light polarization).  The optical qubit of \Ybc is addressed with a 435-nm laser (blue arrow) and, together with a 935-nm repumper (brown line in the inset), is used for sympathetic cooling. Insets: simplified level schemes for \Yb and \Ybc. ({\bf B}) Donor (red) and acceptor (blue) surfaces defined by Eq. \eqref{eq_H} with parameters $(V_x,g,\Delta E)=(0.06,1.6,0.6)\omega$ shown as a function of the reaction coordinate $y$ with their respective noninteracting harmonic wavefunctions. The bath is represented by vibrational modes with a finite linewidth $\gamma$. The color hue reflects the weights of the spin population at each position $y$. ({\bf C}) Donor population dynamics governed by unitary (red circles) and dissipative (blue triangles) processes with $(V_x,g,\Delta E)=(0.18,1,1)\omega$ compared to the numerical results: {The dashed lines with $\gamma=0$ (red) and $\gamma=0.014\omega$ (blue) are calculated from} Eq. \eqref{eq_master}{, whereas their} corresponding solid lines also include spin decoherence ($\gamma_{z}=0.0013 \omega$) and motional dephasing ($\gamma_{m}=0.0013\omega$) (see section S3). Error bars are the statistical SEM.
    }
    \label{fig1}
\end{figure*}

An effective model that describes ET is the celebrated spin-boson model \cite{Leggett1987}. Here, the electronic degrees of freedom are mapped onto a two-level system coupled to a bath of harmonic vibrations encoded in a collection of bosonic modes. This model involves one two-level system, encoding the electron donor and acceptor states, and a reaction coordinate encoded in a single bosonic mode, which is, in turn, itself coupled to a continuous bath of harmonic oscillators \cite{Garg1985, Wolynes1987}. Despite its simplicity, this model allows experimental access to paradigmatic ET regimes by measuring the real-time dynamics of the two-level system and extracting the transfer rate as a function of its coupling to the bosonic mode, the electronic donor-acceptor coupling, their energy difference, and the relaxation rate. The central system is described by the following Hamiltonian \cite{Garg1985, Schlawin2021,MacDonnel2021}, which is a variant of the Rabi model \cite{Lv2018} in quantum optics $(\hbar=1)$

\begin{equation}
    H_{\rm s}= \frac{\Delta E}{2}\sigma_z + V_x \sigma_x + \frac{g}{2} \sigma_z (a^\dagger + a) + \omega a^\dagger a,
    \label{eq_H}
\end{equation}
where $\sigma_{x,z}$ are the Pauli matrices and $a^\dagger (a)$ is the creation (annihilation) operator of the bosonic mode at frequency $\omega$. The reaction coordinate is expressed in terms of the position operator as $y=y_0(a^\dagger + a)/2$, with $y_0=\sqrt{1/2 m\omega}$ and $m$ being the particle mass.
In this model, when $V_x=0$, the energy spectrum is described by two harmonic wells assigned to the donor and acceptor states, $\ket{D}\equiv\ket{\uparrow}_z$ and $\ket{A}\equiv\ket{\downarrow}_z$ separated by a relative energy shift $\Delta E$ (aka exothermicity).
The electronic coupling $V_x$ mixes the states associated with the donor and acceptor surfaces. The spin-boson coupling $g$ displaces the two coupled surfaces along the reaction coordinate, as shown in Fig. \ref{fig1}B. In ET, this is akin to the nuclear coupling that gives rise to the activation energy of a typical ET reaction, which is the core of the Marcus theory \cite{Marcus1985} in chemistry and polaron theory in solid state physics \cite{Lee1953}.

Crucially, the full ET Hamiltonian $H_{ET}=H_{\rm s} + H_{\rm b} + H_{\rm sb}$ must also include bath degrees of freedom $H_{\rm b}$, generally modeled as a large collection of harmonic oscillators, and a linear coupling $H_{\rm sb}$ between the bath and the system's bosonic degree of freedom  \cite{Garg1985}. 
The bath correlation functions and their effect on the system can be described by a continuous spectral density function $J(\omega)$. One way to create an analog for the structured bath spectral densities of biological environments using trapped ions is to use multiple phononic modes naturally hosted in an ion crystal \cite{Clos2016,wang2024simulating}. Here, we take a different approach by exploiting the fact that, under certain conditions, a harmonic environment with a continuous spectral density can be obtained by cooling a spectator ion \cite{Lemmer2018}. In section S6, we prove that sympathetic cooling can effectively simulate an Ohmic spectral density $J(\omega)\sim\omega$, a common choice in the ET literature. 
The cooling process can be described by a master equation in terms of Lindbladian superoperators $\mathcal{L}_{c}[\rho]$, where $c$ is a generic jump operator
\begin{eqnarray}
    \frac{\partial\rho}{\partial t}&=&-i[H_{\rm s},\rho] + \gamma (\bar{n}+1)\mathcal{L}_{a}[\rho] + \gamma \bar{n} \mathcal{L}_{a^\dagger}[\rho],
    \label{eq_master}\\
    \mathcal{L}_{c}[\rho]&=&
    c\rho c^\dagger - \frac{1}{2}\{c^\dagger c,\rho\}.
    \label{eq_Lind}
\end{eqnarray}
Here, $\rho$ is the density matrix of the spin-boson system, $\gamma$ is the motional relaxation rate, and $\bar{n}$ is the phonon population determined by the temperature of the bath $k_B T= \omega/\log(1 + 1/\bar{n})$.

The dynamics of the spin and the bosonic observables predicted by Eq. \eqref{eq_master} are essentially indistinguishable from those of the system in Eq. \eqref{eq_H} in contact with an Ohmic bath, provided that the damping is weak ($\gamma\ll \omega$) and the bath thermal energy is larger than the relaxation rate ($\gamma\beta\ll 1$, with $\beta=1/k_B T$) \cite{Lemmer2018}. As shown in the following, these conditions can be realized experimentally with a trapped-ion system, where the dynamics is determined by five parameters $(\omega, \Delta E, V_x, g,$ and $\gamma)$ that can all be tuned independently. Notably, all the timescales associated with these parameters are faster than the spin and motional decoherence associated with experimental imperfections (see Fig. \ref{fig1}C and section S3), allowing the full characterization of both the transient dynamics and the steady state of the system under Eq. \eqref{eq_master}.

The experimental setup consists of one \Yb ion and one \Ybc ion confined in a linear Paul trap. The two-level system is encoded in the two hyperfine clock states of the \Yb ground-state qubit, $\ket{^2S_{1/2}, F=1,m_F=0}\equiv\ket{\uparrow}_z$ and $\ket{^2S_{1/2}, F=0,m_F=0}\equiv\ket{\downarrow}_z$, separated by a frequency of $\omega_\text{hf}/2\pi = 12.642$ GHz (see Fig. \ref{fig1}A). The bosonic mode in Eq. \eqref{eq_H} is encoded in the radial tilt collective mode at frequency $\omega_{\text{tilt}}/2\pi=3.207$ MHz (see Materials and Methods). 

We engineer $H_{\rm s}$ in Eq. \eqref{eq_H} in a driven rotating frame: Two $\pi/2$ pulses are used to map the $z$ spin basis of Eq. \eqref{eq_H} onto the $y$ basis. In this configuration, two laser tones resonant with the qubit frequency realize the $\Delta E$ and $V_x$ terms. Two additional laser tones at frequencies $\pm\mu=\pm(\omega_\text{tilt}+\delta)$ from the qubit resonance realize the spin-phonon coupling and the harmonic terms in Eq. \eqref{eq_H}, where $\delta\equiv -\omega$ is the detuning with respect to the tilt mode \cite{Schneider_2012}. All the terms in $H_{\rm s}$ are engineered using a 355-nm pulsed laser addressing the \Yb ground-state qubit via stimulated Raman transitions (see Fig. \ref{fig1}A and section S1).

Simulating an independently tunable bath dissipation is achieved by driving the narrow transition from the ground state $\ket{g}\equiv\ket{^2S_{1/2}}$ to the optical metastable state $\ket{o}\equiv\ket{^2D_{3/2}}$ of a \Ybc ion. Two tones of a 435-nm laser combined with a 935-nm repumper are used to perform sympathetic cooling \cite{Cetina2022} on the tilt mode with a cooling rate $\gamma/2\pi$, which is tunable over the 50- to 500-Hz range (see Fig. \ref{fig1}A). This setting is well suited to achieve efficient sympathetic cooling because the fractional mass imbalance of the two ions is very small, and the $\ket{g}\rightarrow\ket{o}$ transition linewidth allows for large Rabi frequencies at modest laser power while providing negligible crosstalk with the qubit states of \Yb.

The experimental protocol (see Fig. \ref{figsequence}) consists of the following steps: (i) After Doppler cooling, Raman-resolved sideband cooling is applied to both the radial center-of-mass and tilt modes. The resulting initial tilt mode phonon population is in the ${\bar{n}_0}\sim(0.1$ to $0.3)$ range, which is comparable to $\bar{n}$ defined in Eq. \eqref{eq_master} and characterized independently by measuring the phonon steady state of the purely dissipative evolution without coherent driving (see fig. S1D). (ii) Then, by applying a $\pi/2$ pulse followed by a displacement operator $\mathcal{D}(-g/2\omega)$, we initialize the system in the donor vibronic state $\ket{D}\bra{D}\otimes \rho_{-}$, where $\rho_{-}=\sum_n e^{- n\omega/k_B T}\ket{n_-}\bra{n_-}$ is a thermal state with temperature $k_B T\approx \omega/\log(1 + 1/\bar{n})$ and $\ket{n_{\pm}}=\mathcal{D}(\pm g/2\omega)\ket{n}$ are displaced Fock states. (iii) We simultaneously apply the laser tones to generate the ET dynamics described by Eq. \eqref{eq_master}. All the parameters that determine the unitary and the dissipative evolutions are calibrated independently (see section S2). (iv) At the end of the evolution, after a final $\pi/2$ pulse, we use state-dependent fluorescence to measure the probability of the system being in the donor state $P_D=(\langle\sigma_z\rangle + 1)/2$ or the average phonon population $\langle n \rangle$ of the tilt mode.

\begin{figure}[t!]
    \centering
    \includegraphics[width=\linewidth]{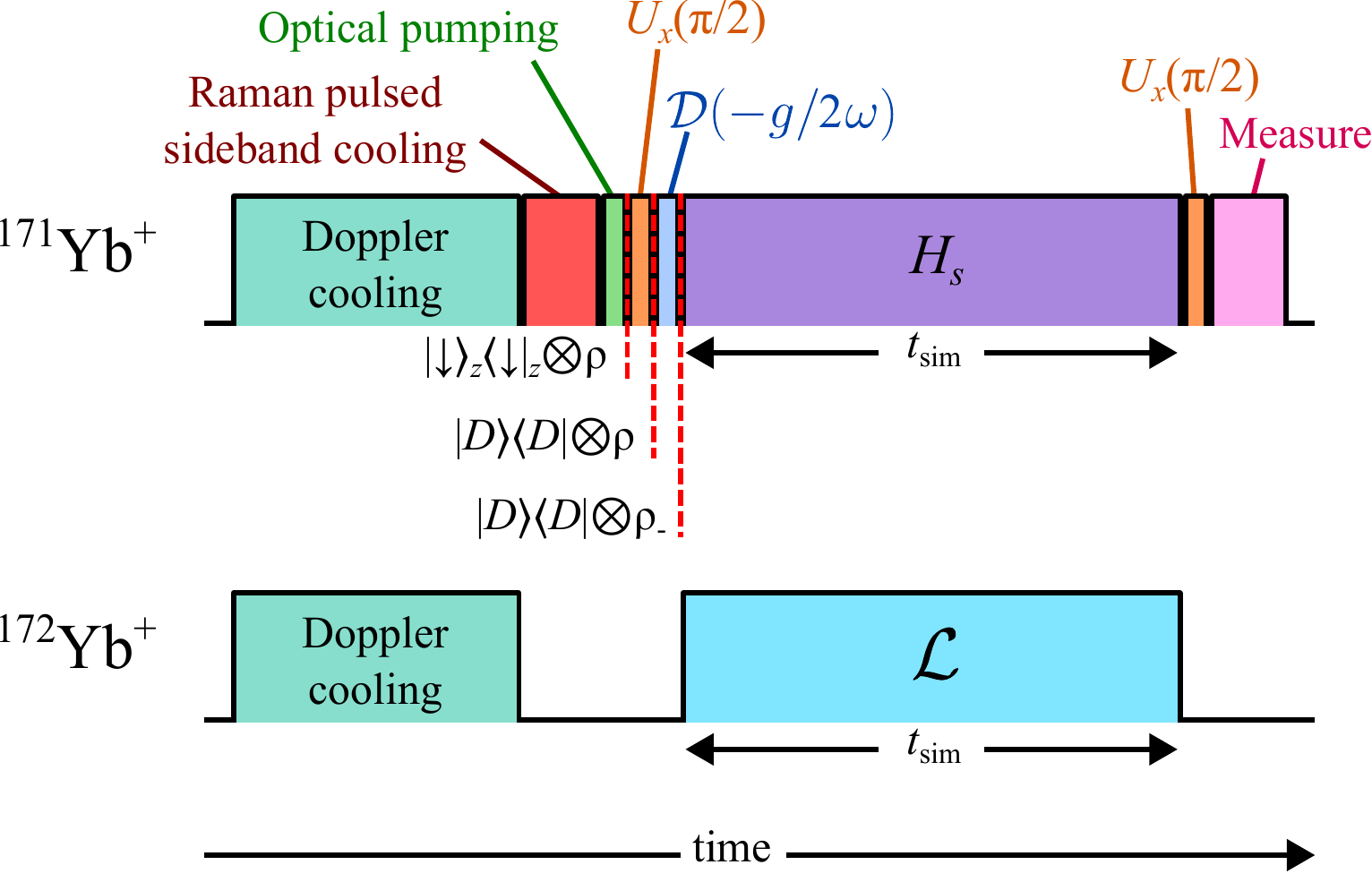}
    \caption{
    {\bf Experimental protocol.} \\After Doppler cooling, Raman sideband cooling, and optical pumping, the initial vibronic state $\ket{D}\bra{D}\otimes \rho_{-}$ is prepared by a $\pi/2$ pulse along the $x$ axis and by displacing the motional state via a spin-dependent force. Before the final measurement, another $\pi/2$ pulse along the $x$ axis rotates the final spin state back into the qubit basis.}
    \label{figsequence}
\end{figure}

The average number of phonons $\bar{n}$ in the (0.1 to 0.3) range fulfills the condition $k_B T \lesssim \omega$ while making sure that the constraint $\gamma\ll k_B T$ is also satisfied. In this highly quantum regime, the transfer is dominated by the discrete level structure of the vibrational mode, and the temperature has a limited effect on the transfer rate. This corresponds to the low-temperature, tunneling-dominated regime of ET.

\begin{figure*}[t!]
    \centering
    \includegraphics[width=1\linewidth]{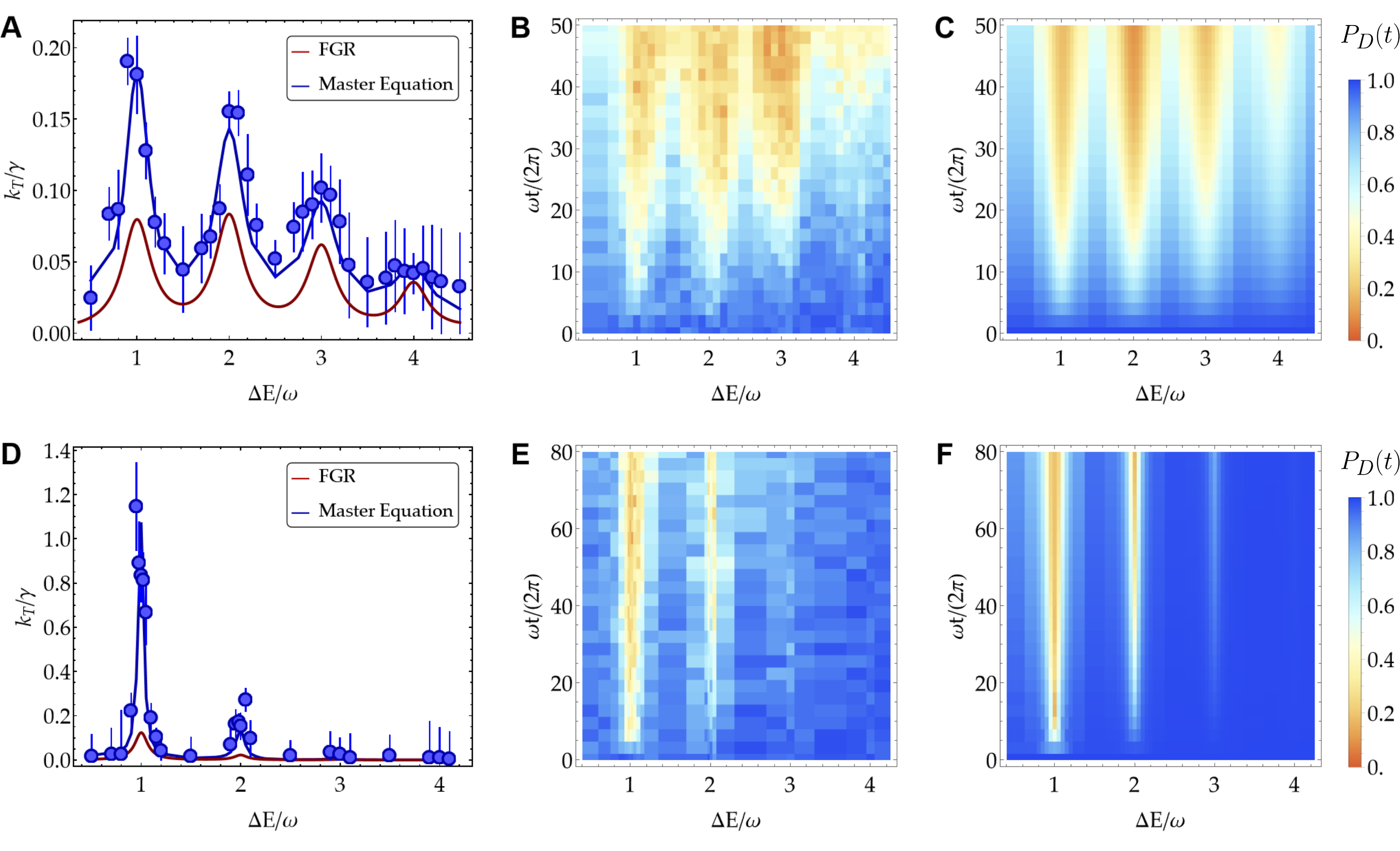}
\caption{{\bf Nonadiabatic transfer regime.} \\({\bf A}) Transfer rate $k_T$ in units of the relaxation rate $\gamma$ as a function of the donor-acceptor energy gap $\Delta E$ for $(V_x, g, \gamma)=(0.056,1.4, 0.06)\omega$. The blue points result from an exponential fit of the measured $P_D(t)$ dynamics, with the error bars being the standard errors of the fit. The dark blue solid curve is obtained from the fit of the dynamics predicted by Eq. \eqref{eq_master}. The FGR prediction (dark red solid line) is calculated using Eq. \eqref{eq_kFermi}. ({\bf B} and {\bf C}) Experimental (B) and numerical (C) density plots of the time-resolved dynamics of $P_D(t)$ as a function of both $\Delta E$ and the number of vibrational oscillations $\omega t/2\pi$. The detuning from the tilt mode is set to $\delta/(2\pi)=-5~\,\rm kHz$, and the numerical results include a motional dephasing of $\gamma_m = 0.001\omega$. ({\bf D}) Transfer rate $k_T$ in units of the relaxation rate $\gamma$ as a function of the donor-acceptor energy gap $\Delta E$ for $(V_x, g, \gamma)= (0.046, 0.521, 0.025)\omega$. 
({\bf E} and {\bf F}) Experimental (E) and numerical (F) density plots of the time-resolved dynamics of $P_D(t)$ as a function of both $\Delta E$ and the number of vibrational oscillations $\omega t/2\pi$, with $\delta/(2\pi)=-10~\,\rm kHz$. The numerical results include a motional dephasing $\gamma_m = 0.0005\omega$ (see section S3).
}
\label{fig2}
\end{figure*}

A crucial parameter for the ET dynamics is the Marcus reorganization energy $\lambda=g^2/\omega$, which is the amount of energy required to displace a wave packet by $g/\omega$ from the center of the donor surface without transferring to the acceptor surface (see Fig. \ref{fig1}B). The reorganization energy, in turn, determines the classical activation energy $U=(\Delta E + \lambda)^2/4\lambda$, which is the barrier a wavepacket localized in the donor surface would have to overcome to enter the acceptor surface when the electronic coupling $V_x$ is negligible. 

We individuate and investigate two regimes \cite{Schlawin2021}: a nonadiabatic and a strongly adiabatic transfer regime. In the former, the electronic coupling $V_x$ is a small perturbation with respect to the other energy scales in the Hamiltonian in Eq. \eqref{eq_H} and is comparable or smaller than the relaxation rate $(V_x\lesssim\gamma)$. When $V_x$ is also much less than $\lambda/4$, namely the activation energy at $\Delta E=0$, the bosonic wave packet is largely localized on either the donor or the acceptor surface, and the ET can be described by the Fermi golden rule (FGR) leading to characteristic isolated peaks in the transfer rate spectrum. Conversely, in the strongly adiabatic regime, the electronic coupling becomes comparable with the activation energy ($V_x \sim \lambda/4$) and greater than the relaxation rate ($V_x>\gamma$), changing the shapes of the donor and acceptor surfaces. In this regime, the transfer rate is less sensitive to the electronic coupling $V_x$ and cannot be predicted by the FGR. Increasing $V_x$ lowers the barrier, and the eigenmodes of Hamiltonian in Eq. \eqref{eq_H} become closer to delocalized wave packets on the two nonadiabatic surfaces. In this case, one can observe significant oscillations between the donor and acceptor states before the steady state is reached (see, for example, Fig. \ref{fig1}C). This regime is realized in a type II or type III mixed valence compound \cite{Demadis2001}. We note that the adiabatic and non-adiabatic regimes are sometimes also called ``coherent'' and ``incoherent,'' respectively. However, in this work, we chose the terminology used in chemical kinetics.




\subsection*{Nonadiabatic regime}

In the nonadiabatic, low-temperature regime, the transfer is dominated by the vibrational mode structure: Both the unitary and dissipative dynamics are frozen unless the donor-acceptor energy difference nearly matches the vibrational energy at $\Delta E = \ell \omega$, with $\ell$ being an integer greater than zero. This vibrationally assisted dynamics \cite{Gorman2018} results in well-resolved resonances (see Fig. \ref{fig2}). 
Deep in the nonadiabatic regime, when $|V_x|\ll\lambda/4$, the eigenstates of the Hamiltonian $H_{\rm s}$ in Eq. \eqref{eq_H} are close to uncoupled donor and acceptor vibronic states represented in Fig. \ref{fig1}B, namely $\ket{D}\ket{n_-}$ and $\ket{A}\ket{n_+}$, respectively. In this case, the $V_x\sigma_x$ term can be treated as a perturbation to the Hamiltonian $H_0=H_{\rm s}-V_x\sigma_x$. As a result, the transfer undergoes resonant transitions between the uncoupled donor and acceptor vibronic states, following the FGR \cite{Leggett1987, Skourtis_1992, Schlawin2021}
\begin{equation}
    k_T=2\pi|V_x|^2 \sum_{n_-,n_+} p_{n_-} \text{FC}_{n_-,n_+}\delta(E_{D,n_-}-E_{A,n_+}),
    \label{eq_kFermi}
\end{equation}
where $p_{n_-}$ is the initial phonon populations in the donor state, and $\text{FC}_{n_-,n_+}=|\langle n_- \ket{n_+}|^2$ is the Franck-Condon factor, namely, the overlap between the two displaced Fock wave functions. A larger displacement $g/\omega$ along the reaction coordinate leads to more vibrational states with a non-negligible overlap, therefore increasing the number of observable transfer resonances.

\begin{figure*}[t!]
    \centering
    \includegraphics[width=1\linewidth]{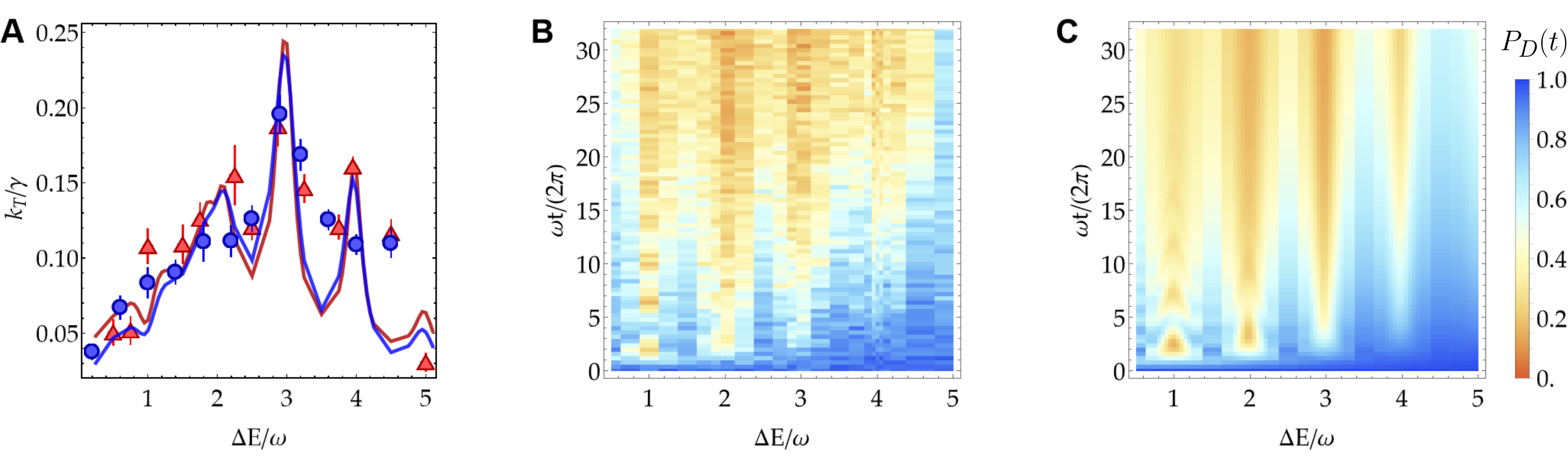}
    \caption{{\bf Adiabatic transfer regime.} \\({\bf A}) Transfer rate $k_T$ measured with $(V_x,g,\gamma)=(0.18, 0.95, 0.020)\omega$  (red triangles) and $(V_x,g,\gamma)=(0.21, 1.08, 0.038)\omega$ (blue circles). The solid curves are the transfer rates calculated from Eq. \eqref{eq_master} using the definition in Eq. \eqref{eq_lifetime_kT} and including spin decoherence ($\gamma_{z}=0.0025 \omega$) and motional dephasing ($\gamma_{m}=0.0013\omega$).
    The transfer rates overlap when scaled in units of the relaxation rate $\gamma$. The error bars are calculated using bootstrapping (see Materials and Methods).
    ({\bf B}) Experimental donor population evolution $P_D(t)$ versus energy gap $\Delta E$ and the number of vibrational oscillations $\omega t/2\pi$ with the same parameters as the red triangles in {(A)}. Here, the detuning from the tilt mode is set to $\delta/2\pi=-4~\,\rm kHz$.
    ({\bf C}) Corresponding numerical results with the same parameters as {(B)}. 
    }
    \label{fig3}
\end{figure*}

In this regime, the effect of the bath can be taken into account by replacing the delta functions in Eq. \eqref{eq_kFermi} with normalized Lorentzian distributions with full width at half maximum $\gamma$, namely, $\delta(E_{D,n_-}-E_{A,n_+})\rightarrow (\gamma/2\pi)/\left[(E_{D,n_-}-E_{A,n_+})^2 + (\gamma^2/4)\right]$. 

In Figs. \ref{fig2}A, we show the transfer rates extracted from the dynamics of the donor population $P_D(t)$, shown in Fig. \ref{fig2}B (experimental data) and \ref{fig2}C (theory) as density plots as a function of $\Delta E$ and the number of vibrational oscillations $\omega t/2\pi$. The transfer rates extracted from an exponential decay fit of $P_D(t)$ agree with the numerical predictions from the Lindblad master equation in Eq. \eqref{eq_master}, exhibiting distinct peaks at $\Delta E=\ell \omega$. In Fig. \ref{fig2}(A to C), the chosen parameters place the system in the nonadiabatic regime ($V_x=0.056\omega$ and $\lambda/4=0.49\omega$), which is confirmed by the qualitative agreement between the FGR prediction (dark red solid line), the experimental results, and the exact theory. Here, because $g=1.4\omega$, we can observe transfer resonances involving vibrational states up to $n=4$ within our experimental resolution (see section S3).

In Fig. \ref{fig2}(D to F), we decrease the spin-motion coupling to $g=0.521\omega$ and the motional relaxation rate to $\gamma=0.025\omega$ while keeping the values of the other parameters approximately the same as in Fig. \ref{fig2}(A to C). In this case, by lowering the spin-motion coupling strength, the Franck-Condon coefficients $\text{FC}_{n_-,n_+}$ are greater for smaller $n$ compared to Fig. \ref{fig2}(A to C). Therefore, fewer vibrational excitations are involved in the transfer through the effective vibronic coupling strength $V_x\sqrt{\text{FC}_{n_-,n_+}}$, resulting in the reduction in the number of observed resonances compared to Fig. \ref{fig2}(A to C). In addition, the data in Fig. \ref{fig2}(D to F) show that the FGR predictions in Eq. \eqref{eq_kFermi} further underestimate both the experimental and numerical results as the system is approaching the nonperturbative regime.  Lastly, the decrease in the motional relaxation rate makes the width of the resonances sharper across the spectrum, confirming its connection with the broadening of the vibrational modes.

\subsection*{Strongly adiabatic regime}

When the electronic coupling $V_x$ is comparable to the activation barrier $\lambda/4$ and larger than the relaxation rate $\gamma$, the dynamics cannot be simply described in terms of weakly coupled wave function localized on the donor and acceptor site. 
In this regime, the population evolution features an initial coherent oscillation between the donor and acceptor states before the eventual equilibration in the acceptor state, as shown in Fig. \ref{fig3}(B and C). Here, the density plots of the experimental and theoretical $P_D(t)$ are plotted as a function of $\Delta E$, showing good agreement.
In this regime, the evolution cannot be fitted with an exponential function as in the nonadiabatic case. Therefore, to extract the effective transfer rate, we use the inverse lifetime of the donor population as proposed in refs. \cite{Skourtis_1992, Schlawin2021}
\begin{equation}
    k_T^{-1}=\frac{\int t P_D(t)dt}{\int P_D(t)dt}.
    \label{eq_lifetime_kT}
\end{equation}
In Fig. \ref{fig3}A, 
the transfer rates are extracted using Eq. \eqref{eq_lifetime_kT} by interpolating and integrating both the experimental data and the numerical results (see Materials and Methods).
We show the transfer rates extracted from the data for two sets of parameters that have nearly equal spin-phonon coupling $g$ and electronic coupling $V_x$ but different relaxation rates $\gamma$. We report the results in units of $\gamma$, showing that the transfer rate is proportional to the relaxation rate ($k_T\propto\gamma$). In this regime, $\gamma$ becomes the limiting factor for the rate at which the donor state population irreversibly transfers into the acceptor state. This result can be explained intuitively by considering a simplified two-vibronic-state model \cite{Schlawin2021} consisting of the initial donor state $\ket{D}\ket{0_-}$ and a single acceptor vibronic state $\ket{A}\ket{\nu_+}$ with the coupling strength $V_x\sqrt{\text{FC}_{0_-,\nu_+}}$ and a decay rate $\nu\gamma$, where $\nu=\Delta E/\omega$. In this simplified case, the transfer rate in Eq. \eqref{eq_lifetime_kT} can be evaluated analytically as
\begin{equation}
    k_T^{0,\nu}=\nu\gamma\frac{1+\left(\frac{\nu\gamma}{V_x\sqrt{\text{FC}_{0_-,\nu_+}}}\right)^2}{1+\frac{1}{2}\left(\frac{\nu\gamma}{V_x\sqrt{\text{FC}_{0_-,\nu_+}}}\right)^4}.
    \label{eq_adiabatic_kT}
\end{equation}
When $V_x\sqrt{\text{FC}_{0_-,\nu_+}}\gg\gamma$, $k_T^{0,\nu}\approx\nu\gamma$. Although this approximation rightly predicts the proportionality between $k_T$ and $\gamma$ in the strongly adiabatic regime, it fails to accurately predict the transfer rates when more than one vibronic acceptor state is involved.

A few comments are in order: 
(i) For $\Delta E<2\omega$, the transfer rate $k_T$ does not exhibit distinct resonances as opposed to the transfer rate in the nonadiabatic regime.
(ii) For $\Delta E>2\omega$, the characteristic peaked structure of the nonadiabatic regime is recovered, which can be explained by the localization of the initial state in the upper hybridized surface, as suggested by ref. \cite{Schlawin2021}.
(iii) For $\Delta E>3\omega$, the envelope of the transfer rate shows a decrease as a function of $\Delta E$. This is sometimes called the ``inverted regime" of ET, where, at both high and low temperatures, the reaction counterintuitively becomes slower despite the transfer becoming more exothermic. This can be explained by the decreasing Franck-Condon factor ${\rm FC}_{n_-,n_+}$ as a function of $\Delta E$ and can also be observed in the nonadiabatic regime (see Fig. \ref{fig2}).

\begin{figure}[t!]
    \centering
    \includegraphics[width=\linewidth]{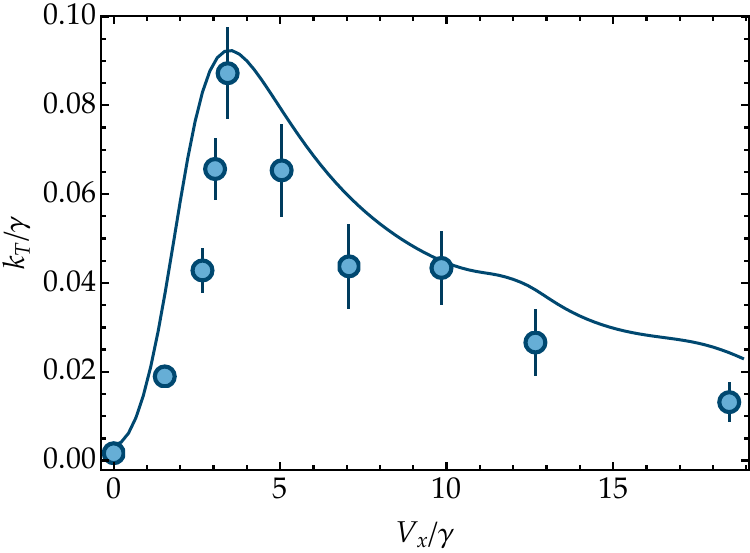}
    \caption{
    {\bf Optimal transfer.} \\Transfer rate $k_T$ as a function of $V_x/\gamma$, with $(\Delta E,g,\gamma)=(2,0.80,0.11)\omega$ and detuning $\delta/2\pi = -4$ kHz. The numerical results (solid curve) include spin decoherence ($\gamma_{z}=0.0013 \omega$) and motional dephasing ($\gamma_{m}=0.0013\omega$). The optimal transfer is located at $V_x/\gamma\sim3.3$, in agreement with the theoretical prediction of Eq. \eqref{eq_master}. Error bars are calculated using bootstrapping (see Materials and Methods).}
    \label{fig4}
\end{figure}

\subsection*{Optimal transfer}

When $\Delta E$ is set on a resonance, sweeping $V_x/\gamma$ allows one to pinpoint an optimal transfer regime \cite{Skourtis_1992}.  
In Fig. \ref{fig4}, we report the transfer rate measured as a function of $V_x/\gamma$, setting $\Delta E= 2 \omega$. The data exhibit a distinct optimal transfer rate at $V_x/\gamma\sim3.3$, in good agreement with the numerical predictions based on Eq. \eqref{eq_master}. It is worth noting that, for small $V_x/\gamma$, the transfer rate varies quadratically as predicted by Eq. \eqref{eq_kFermi}. Beyond the optimum, the transfer rate is less sensitive to $V_x/\gamma$. This robustness has been suggested to be important for fast transfer in photosynthetic complexes \cite{Onuchic1988, Skourtis_1992}. In particular, the presence of an optimal relaxation rate underscores the crucial role of dephasing in transport phenomena that was previously pointed out in solid-state  \cite{Logan1987} and atomic systems \cite{Maier2019}, as well as in biomolecules \cite{Plenio2008, Wolynes2009, Rebentrost2009, Chin2010}.

\section*{Discussion}

Our experiment demonstrates the remarkable flexibility of the trapped-ion platform to perform direct analog quantum simulations of models relevant to chemical physics, including an engineered environment.
These simulations are performed through careful tuning of both the Hamiltonian of the trapped-ion system and its engineered reservoir by using seven simultaneous laser tones and two different atomic species.
This toolbox allowed us to investigate relevant regimes of a paradigmatic ET model with tunable dissipation at low temperatures, where the interplay of quantum effects and interactions with the environment is crucial in determining the dynamics.
The observed time-resolved dynamics of the donor-acceptor population and the measured transfer rate in both the nonadiabatic and adiabatic regimes agree with the numerics with independently calibrated parameters and identify an optimal transfer regime that has been suggested to be relevant for ET in photosynthetic complexes \cite{Skourtis_1992}. 

{
Our experiment opens up new avenues for simulating condensed-phase chemical quantum dynamics. The trapped-ion simulator allows native encoding of the bosonic degrees of freedom and their tunable dissipation without the need for digitization, leading to linear scaling with both the number of electronic states and that of bosonic modes. In this context, existing classical numerical methods used to solve these models are more computationally expensive when the reorganization energy is of the same order or larger than the electronic coupling ($\lambda \gtrsim V_x$) \cite{Somoza2019, Kang2024}. 
To access such a parameter range, it is necessary to experimentally realize nonperturbative spin-phonon couplings $g\gtrsim\omega$. Crucially, the approach used here based on sympathetic cooling gives rise to the dynamics of the corresponding spin-boson model with a Lorentzian spectral density at all orders in $g$ under the assumptions used in this work ($\gamma\ll\omega,\gamma\beta\ll1$) \cite{Lemmer2018, Tamascelli2018}. 
Therefore, this approach will enable the realization of structured spectral density functions \cite{Lemmer2018} and the simulation of colored baths and non-Markovian dynamics \cite{deVega2017, Harrington2022} by using multiple ions as coolants to control the individual cooling rates and the temperature of multiple bosonic modes.

To investigate the role of coherence and Frenkel-type exciton delocalization \cite{Jang2018} in the energy transfer processes in biomolecules and photosynthetic complexes \cite{Ishizaki2009, Mattioni2021}, a necessary extension is the encoding of multiple electronic excited states. This can be achieved using more than two atomic levels (a qudit) coupled to phonons \cite{MacDonnel2021} provided by the ion crystal. Alternatively, multiple electronic states (sites) can be physically mapped to qubit ions and individually addressed to tailor their energy landscape and their individual couplings to the phonon bath. At the same time, the site qubits will have to be connected via a long-range spin-hopping Hamiltonian that can be realized with Mølmer-Sørensen Ising interactions \cite{monroe2021programmable}. 
}
In addition, the trapped-ion platform naturally offers the possibility to include tunable anharmonic couplings among different bosonic modes \cite{Ding2017} that can be used to study the effects of anharmonicity on energy transfer \cite{Zhang2023pnas}, a crucial but often overlooked feature of realistic molecular systems that hinders the applicability of existing numerical methods.

The native long-range character of the spin-spin interactions and the presence of collective bosonic modes with tunable dissipation and anharmonic couplings will allow the simulation of out-of-equilibrium chemical dynamics that are challenging to address with classical methods. 
%
%
Our experiment is therefore a stepping-stone toward the use of quantum devices to provide insights into open questions in chemical and biological physics and to shed light on the underlying principles of biochemical processes.

During the preparation of this manuscript, we became
aware of a complementary work \cite{sun2024quantumsimulationspinbosonmodels}, which simulates the dephased spin-boson model using randomized unitary spin-dependent forces.

\section*{Materials and Methods}

\subsection*{Experimental system}
The experimental system is based on a blade trap, where each blade features five segmented electrodes. We mounted the gold-coated fused silica blades on an Alumina holder. Alumina is chosen for its high thermal conductivity and low outgassing rate. The blades are positioned in a $60^\text{o}/30^\text{o}$ angle configuration to enable high optical access along the vertical direction for high-resolution imaging [0.6 numerical aperture (NA)] and along the in-plane direction orthogonal to the trap axis (0.3 NA). This configuration also breaks rotational symmetry, which allows for well-defined trap principal axes. Each electrode is biased via a gold fuzz button, which is, in turn, connected to a Kapton-insulated wire via customized Macor holders. To shunt the radio frequency (rf) pickup voltages on the static dc blades, we use ultrahigh vacuum-compatible silver-filled epoxy to glue 800-pF capacitors to each static segment on one side and wire bond the other side to a ground strip present on the blades.
We use a helical resonator with a resonant frequency of 27.9-MHz and a quality factor $Q=198$ to drive the rf blades, achieving a radial center-of-mass trap frequency of $3.363$-MHz at $V_{\rm peak}=420$-V.
The heating rate on the radial center-of-mass mode is measured to be 0.4 quanta/ms, whereas the tilt mode features a lower heating rate ($\dot{n}\sim 0.03$ quanta/ms).

A 370-nm laser red detuned from the ${^2S}_{1/2}\rightarrow{^2P}_{1/2}$ transition passing through 3.704- and 14.748-GHz electro-optic modulators is used to produce Doppler cooling light for both isotopes \Yb and \Ybc. This beam is placed in-plane at $45^\text{o}$ with respect to the ion chain for projection along all three trap principal axes. In addition, two axial 370-nm beams are used for detection and optical pumping of \Yb. They are also superimposed with two 935-nm repumper beams for both Yb$^+$ isotopes.

A pulsed 355-nm laser is used to resonantly address the \Yb ground-state qubit via two-photon Raman processes. The same laser is used to generate the spin-phonon coupling. The counterpropagating Raman beams have elliptical shapes with vertical and horizontal waists $w_{z}=5\,{\rm\mu m}$ and $w_{x}=150\,{\rm\mu m}$ and are in lin$\perp$lin polarization configuration to maximize the coupling between the two hyperfine clock states.

A 435-nm diode laser locked to an ultralow expansion cavity is used to address the ${^2S}_{1/2}\rightarrow{^2D}_{3/2}$ transition (or $\ket{g}\rightarrow\ket{o}$) in \Ybc \cite{Tamm2014, Allcock2021}. The beam is aligned at $45^o$ with respect to the magnetic field and horizontally polarized to maximize the coupling to the two $\Delta m_j=0$ transitions ($m_j=\pm1/2\rightarrow m_{j'}=\pm1/2$) separated by $8.23$ MHz. The cooling is achieved by continuously driving the red sideband of $m_j=\pm1/2\rightarrow m_{j'}=\pm1/2$ transitions while also using a 935-nm repumper laser that allows the transition between $\ket{o}$ and ${^3D[3/2]}_{1/2}\equiv\ket{e}$. Two tones on the 935 nm laser separated by 113-MHz address both \Yb and \Ybc. To avoid optical pumping into either of $m_j=\pm1/2$ ground states during continuous sideband cooling, we use two laser tones on the 435-nm laser to address both the $m_j=\pm1/2\rightarrow m_{j'}=\pm1/2$ transitions simultaneously. The effective cooling rate is highly dependent on the power of the 935-nm laser, and it is the main turning knob to tune the cooling rate $\gamma$. 

\subsection*{Experimental sequence}
\label{app_sequence}

The experimental procedure is summarized in Fig. \ref{figsequence}. Our setup consists of a \Yb ion acting as the qubit and a \Ybc ion acting as the coolant. 
Initially, we use the standard Doppler cooling technique on both ions to prepare the temperature of the trapped dual-species chain near the Doppler limit. 
We then perform the Raman-resolved sideband cooling protocol on the radial center-of-mass and tilt modes, followed by an optical pumping pulse, to prepare the system in $\ket{\downarrow}_z\bra{\downarrow}_z\otimes \rho$, where $\rho=\sum_n e^{- n\omega/k_B T}\ket{n}\bra{n}$ is the thermal phonon density matrix of the tilt mode and $k_B T= \omega/\log(1 + 1/\bar{n}_0)$ is the associated temperature. The initial tilt mode average phonon $\bar{n}_0$ is set to range between 0.1 and 0.3, which is similar to the bath temperature $\bar{n}$. To transform the system from the qubit basis $\sigma_z$ to the $\sigma_y$ basis, we apply a global rotation $U_x(\pi/2)=\exp(-i \sigma_x\pi/4)$.
The state of the system becomes $\ket{D}\bra{D}\otimes \rho$, where $\ket{D}\equiv\ket{\uparrow}_y$ here. 
\par We then prepare the motional population from $\rho$ to $\rho_-$ with an optical dipole force from two Raman beatnotes, $\omega_r = \omega_\text{hf}-\mu$ and $\omega_b = \omega_\text{hf}+\mu$, which have the same Rabi coupling strength of $\Omega^\text{displace}=\Omega/2$, with $\eta\Omega=g$ and $\phi_r = \phi_b = \pi$. We point out that this is the same drive that generates the spin-phonon term in Eq. \eqref{eq_H} but with half the Rabi coupling strength. This results in a spin-dependent displacement of the motional state
\begin{equation}
        H_\text{displace}^\text{eff} = \frac{\eta\Omega^\text{displace}}{2}\sigma_y\left(a e^{i\delta t} + a^\dagger e^{-i\delta t}\right),
        \label{eq_Hord}
\end{equation}
where $\delta\equiv\mu - \omega_\text{tilt}$.
Under this operation, the system evolves as $U(t)=\mathcal{D}[\alpha(t)]\ket{\uparrow}_y\bra{\uparrow}_y+\mathcal{D}[-\alpha(t)]\ket{\downarrow}_y\bra{\downarrow}_y\equiv\mathcal{D}[\alpha(t)]\ket{D}\bra{D}+\mathcal{D}[-\alpha(t)]\ket{A}\bra{A}$, where $\mathcal{D}$ is the displacement operator in position-momentum phase space and $\alpha(t) = \alpha_0(1-e^{-i\delta t})$ with $\alpha_0=\eta\Omega^\text{displace}/2\delta=g/4\delta$ \cite{monroe2021programmable}. Hence, the applied pulse duration is $t_\text{displace}=\pi/\delta$ to get the displacement of $\alpha(t_\text{displace})=g/2\delta=-g/2\omega$ onto $\rho$.
\par With the system being in the desired initial state $\ket{D}\bra{D}\otimes \rho_-$, we address the \Yb with the four Raman beatnotes to generate an effective Hamiltonian that maps to the ET unitary model in Eq. \eqref{eq_H} (see section S1).
Simultaneously, we apply the continuous resolved sideband cooling protocol on the \Ybc's narrow linewidth optical transition to sympathetically cool the tilt mode of the system at the rate $\gamma$ and effectively realize an engineered phonon dissipation. By varying the simulation time $t_\text{sim}$, we can measure the time-dependent evolution of the system. Before the measurement, we rotate the system back to the qubit basis with another global rotation $U_x(\pi/2)$.
\par To measure the average spin excitation, we use spin-dependent fluorescence, where only the spin in state $\ket{\uparrow}_z$, now representing $\ket{D}$ after the $\pi/2$ global rotation, scatters photons. 
We use an objective lens with an NA of 0.6 to collect the scattered photons into the photomultiplier tube. The average state discrimination fidelity between $\ket{D}$ and $\ket{A}$ is 99.5$\%$.

Alternatively, we can measure the average phonon excitation $\braket{a^\dagger a}$ by performing an optical pumping pulse to reset the spin state of the system to $\ket{\downarrow}_z$ followed by a resonant Raman blue sideband (BSB) transition drive, $H^\text{BSB}=i(\eta\Omega/2)(a\sigma^- - a^\dagger\sigma^+)$, before the average spin excitation measurement. The phonon-number distribution that represents the diagonal elements of the final phonon density matrix of the system, $\rho_m$, can be extracted by fitting the spin  evolution under the resonant Raman BSB transition drive with
\begin{equation}
        P_{\ket{\uparrow}_z}(t)=\frac{1}{2}\sum_n p(n)\left[1-e^{-\alpha_m t}\cos(\sqrt{n+1}\eta\Omega t)\right],
\end{equation}
where $p(n)$ denotes the phonon-number state population, $\alpha_m$ is a parameter to capture the decoherence rate of the spin-phonon evolution, and $t$ is the drive time \cite{Lv2018}. Hence, we can compute $\braket{a^\dagger a} = \text{Tr}\left(\rho_m a^\dagger a\right)$.

\subsection*{Transfer rate data analysis}

In the nonadiabatic regime, the transfer dynamics can be well described by an exponential decay (see figs.~S3, A and B).
Because of the finite bath temperature $\bar{n}\sim 0.1$ to $0.3$, the spin population transfer is not complete from $\ket{D}$ to $\ket{A}$. Therefore, the transfer rates are extracted from an exponential function with the rates and final populations as the fitting parameters. The uncertainties of the rates are the corresponding standard errors of the fits.

On the other hand, the spin evolutions in the adiabatic regime feature complex oscillatory decays that a simple analytical model cannot describe (see figs. S3, C and D). For this reason, we use the inverse lifetime of the donor population in Eq. \eqref{eq_lifetime_kT} to determine the transfer rates \cite{Skourtis_1992, Schlawin2021}. This definition considers $t \rightarrow \infty$; therefore, there is a correction we need to consider when we use this formula for a finite experimental time. In the case of no electronic coupling, $V_x = 0$, the donor population does not evolve, $P_D(t) = 1$, because it is in an eigenstate of the system. However, Eq. \eqref{eq_lifetime_kT} still evaluates a nonzero transfer rate between $t=0$ and $t = t_\text{sim}$ as $k_0 = \frac{2}{t_\text{sim}}$. This contribution to the transfer rate only goes to zero if one evaluates Eq. \ref{eq_lifetime_kT} for $t\rightarrow\infty$. 
Because $P_D$ reaches the steady state within our experimental resolution in a finite time $t_{\text{sim}}$ ranging from 4 to 10 ms, we calculate the transfer rates by subtracting $k_0$ as
\begin{equation}
    k_T=\frac{\int_{0}^{t_\text{sim}} P_D(t)dt}{\int_{0}^{t_\text{sim}} t P_D(t)dt} - k_0.
    \label{eq_mod_lifetime_kT}
\end{equation}
To numerically evaluate the integrals, we interpolate the evolution $P_D(t)$ data. 
We also use Eq. \eqref{eq_mod_lifetime_kT} to estimate the numerical transfer rates.

\par To estimate the errors of the transfer rate, we follow a resampling procedure. We consider the experimental error of each time step of the $P_D(t)$ measurements as the SD of a normal distribution centered at the mean measured value. We then randomly sample the distributions at each time step, and we estimate the error of the transfer rate by taking the SD of the rates obtained from the resampled datasets by using Eq. \eqref{eq_mod_lifetime_kT}.
The process is repeated for all adiabatic transfer dynamics. 

\nocite{Schneider_2012}
\nocite{monroe2021programmable}
\nocite{JOHANSSON2013}
\nocite{Fluhmann2019encoding}
\nocite{Whitlow2023}
\nocite{Valahu2023}
\nocite{Lv2018}
\nocite{huelga2012non}
\nocite{Petruccione2007}
\nocite{Garg1985}
\nocite{carmichael2013statistical}
\nocite{brandes2004chapter}
\nocite{mohseni2014quantum}
\nocite{nielsen2001quantum}
\nocite{Lemmer2018}

\begin{acknowledgments}
We acknowledge Y. Tanimura for insightful discussions and F. Minganti, D. Fallas-Padilla, and M. Dalmonte for suggestions and careful reading of the manuscript. We acknowledge A. Sheffield for early contribution to the experimental setup.

\noindent\textbf{Funding:} G.P. acknowledges the support of the Welch Foundation Award C-2154, the Office of Naval Research Young Investigator Program (grant no. N00014-22-1-2282), the NSF CAREER Award (grant no. PHY-2144910), the Army Research Office (W911NF22C0012), and the Office of Naval Research (grant no. N00014-23-1-2665). We acknowledge that this material is based on work supported by the U.S Department of Energy, Office of Science, Office of Nuclear Physics under the Early Career Award No. DE-SC0023806. The isotopes used in this research were supplied by the US Department of Energy Isotope Program, managed by the Office of Isotope R$\&$D and Production. H.P. acknowledges support from the NSF under grant no. PHY-2207283. Work at the Center for Theoretical Biological Physics was supported by the NSF (grant no. PHY-2019745). J.N.O. was also supported by the NSF grant no. PHY-2210291. P.G.W. was also supported by the D. R. Bullard-Welch Chair at Rice University (grant no. C0016).

\noindent\textbf{Author contributions:} V.S., M.D.S., A.M., R.Z., and G.P. contributed to the experimental design, construction, data collection, and analysis of this experiment. M.Z., H.P., J.N.O., and P.G.W. contributed to the paper’s conceptualization and supporting theory and numerics. All authors contributed to the writing and revision of the manuscript.

\noindent\textbf{Competing interests:} R.Z. is a cofounder and chief executive officer at TAMOS Inc. G.P. is a cofounder and chief scientist at TAMOS Inc. The other authors declare that they have no competing interests.

\noindent\textbf{Data and materials availability:} All data needed to evaluate the conclusions in the paper are present in the paper and/or the Supplementary Materials. The source data are available in the Zenodo repository (https://zenodo.org/records/13858935).
\end{acknowledgments}

\bibliography{ET}

\section*{Supplementary Materials}

\setcounter{figure}{0}
\renewcommand{\figurename}{Figure}
\renewcommand{\thefigure}{S\arabic{figure}}

\subsection*{S1 Hamiltonian derivation} \label{SM_S1}

In this section, we derive the mapping from the experimental trapped-ion Hamiltonian to the electron transfer model in Eq. \eqref{eq_H} in the main text. When we apply a pair of counterpropagating Raman beams with a wavevector difference of $\vec{k}$, phase difference of $\phi$, and a beatnote frequency at $\omega_L$ on the \Yb trapped qubit in a dual-species chain, the system can be described by ($\hbar = 1$)
\begin{equation}
\begin{split}
H = \frac{\omega_\text{hf}}{2}\sigma_z &+\sum_\nu\omega_\nu a^\dagger_{\nu} a_{\nu} \\
&+ \frac{\Omega}{2}\left(e^{\sum_\nu i\eta_\nu\left(a_\nu +a^\dagger_\nu \right)-i\omega_L t-i\phi}\sigma^+ + \text{h.c.}\right),
\end{split}
\label{eq_Hfull}
\tag{S.1}
\end{equation}
where $\omega_\text{hf}$ is the energy splitting between the two qubit states, $\omega_\nu$ is the $\nu$-th collective motional frequency of the chain associated with the raising (lowering) operator, $a^\dagger_\nu (a_\nu)$, $\Omega$ is the Rabi coupling strength, and $\eta_{\nu} = k\sqrt{1/2m\omega_\nu}b_\nu$ is the Lamb-Dicke parameter, and $m$ is the qubit mass. $b_\nu$ is the normalized motional eigenvector for the \Yb qubit ion in the $\nu=1,2$ modes, namely, the com and tilt modes of the \Yb-\Ybc crystal.
 
By adding and subtracting $\sum_\nu\delta_{\nu} a^\dagger_\nu a_\nu$ to Eq. \eqref{eq_Hfull} and rotating with respect to $\frac{\omega_\text{hf}}{2}\sigma_z + \sum_\nu \mu a_\nu^\dagger a_\nu$, Eq. \eqref{eq_Hfull} is transformed into a resonant interaction frame rotating at $\mu = \omega_L - \omega_\text{hf} \equiv \omega_\nu+\delta_\nu$, where $\delta_\nu$ is the detuning from the $\nu$-th motional mode \cite{Schneider_2012}. In our experiment, $\mu+\omega_\nu\gg|\mu-\omega_\nu|=\delta_\nu$, therefore a rotating-wave approximation (RWA) is justified, and terms that rotate at $\mu+\omega_\nu$ can be neglected. After the RWA,  the Hamiltonian is described by
\begin{equation}
    \begin{split}
        H_I^\text{res} = \frac{\Omega}{2}\left(e^{i\sum_\nu\eta_\nu\left(a_\nu e^{-i \mu t}+a_\nu^\dagger e^{i \mu t}\right)}e^{-i(\omega_L - \omega_\text{hf})t-i\phi}\sigma^+ \right.\\\left.+ \text{h.c.}\right)-\sum_\nu\delta_\nu a^\dagger_\nu a_\nu.
    \end{split}
    \label{eq_interactionH0_2modes}
    \tag{S.2}
\end{equation}
For our setup, the detuning from the tilt mode ($\delta_{\rm tilt}/2\pi\equiv \delta/2\pi$ in the main text) ranges from -4 to -10 kHz, while $\delta_{\rm com}/2\pi\sim -160\,\rm kHz$. Therefore, we can safely neglect the contribution from the com mode, obtaining a single-mode Hamiltonian
\begin{equation}
    \begin{split}
        H_I^\text{res} = \frac{\Omega}{2}\left(e^{i\eta\left(a e^{-i \mu t}+a^\dagger e^{i \mu t}\right)}e^{-i(\omega_L - \omega_\text{hf})t-i\phi}\sigma^+ + \text{h.c.}\right)\\-\delta a^\dagger a.
    \end{split}
    \label{eq_interactionH0}
    \tag{S.3}
\end{equation}
In the experiment, we apply four tones to one of the Raman beams generating four beatnotes at frequencies 
$\omega_r = \omega_\text{hf}-\mu$ with phase $\phi_r$ (red sideband or RSB),
$\omega_b = \omega_\text{hf}+\mu$ with phase $\phi_b$ (blue sideband or BSB), 
$\omega_x = \omega_\text{hf}$ with phase $\phi_x$, and 
$\omega_y = \omega_\text{hf}$ with phase $\phi_y$. Thus, Eq. \eqref{eq_interactionH0} becomes
\begin{equation}
    \begin{split}
        H_I^\text{res} = \sum_{k=r,b,x,y}\frac{\Omega_k}{2}\left(e^{i\eta\left(a e^{-i\mu t}+a^\dagger e^{i\mu t}\right)}e^{-i\left(\omega_k - \omega_\text{hf}\right) t-i\phi_k}\sigma^+\right. \\+ \Big.\text{h.c.}\Big)-\delta a^\dagger a.
    \end{split}
    \label{eq_interactionH4}
    \tag{S.4}
\end{equation}
The first two terms in the summation generate the spin-phonon coupling term in Eq. \eqref{eq_H} in the main text. In the Lamb-Dicke regime, where $\eta\sqrt{\braket{\left(a+a^\dagger\right)^2}}\ll1$, we can expand the two terms with respect to $\eta$ to the first order and apply rotating-wave approximation to neglect off-resonant terms rotating at $\mu$ and $2\mu$. For $\Omega_r = \Omega_b \equiv \Omega$, we obtain the effective spin-phonon Hamiltonian
\begin{equation}
        H_\text{sp} = \frac{\eta\Omega}{2}\left(a e^{i\phi_m} + a^\dagger e^{- i\phi_m}\right)\left(\cos\phi_s\sigma_x+\sin\phi_s\sigma_y\right),
        \tag{S.5}
\end{equation}
where the motional phase $\phi_m \equiv \frac{\phi_b - \phi_r}{2}$ and the spin phase $\phi_s \equiv \frac{\phi_b + \phi_r}{2}-\frac{\pi}{2}$ 
\cite{monroe2021programmable}. We choose $\phi_r = \phi_b = \pi$ for the experiment. Hence, the Hamiltonian is further simplified to
\begin{equation}
        H_\text{sp} = \frac{\eta\Omega}{2}\sigma_y\left(a + a^\dagger \right).
        \label{eq_Heff}
        \tag{S.6}
\end{equation}
The two remaining terms in the summation in Eq. \eqref{eq_interactionH4} follow the same form, differed by only the phase difference $\phi_k$ with $k=x,y$ and, in the Lamb-Dicke regime,
generate
\begin{equation}
        H_k = \frac{\Omega_k}{2}\left(\cos\phi_k\sigma_x+\sin\phi_k\sigma_y\right).
        \label{eq_Hceff}
        \tag{S.7}
\end{equation}
\par By substituting Eq. \eqref{eq_Heff} and Eq. \eqref{eq_Hceff} with $\phi_x = 0$ and $\phi_y = \pi/2$ into Eq. \eqref{eq_interactionH4}, we obtain
\begin{equation}
        H_I^\text{res} = \frac{\Omega_y}{2}\sigma_y + \frac{\Omega_x}{2} \sigma_x + \frac{\eta\Omega}{2} \sigma_y (a^\dagger + a) - \delta a^\dagger a.
        \label{eq_interactionHion}
        \tag{S.8}
\end{equation}

As explained in the main text, to map Eq. \eqref{eq_interactionHion} to the ET model in Eq. \eqref{eq_H} in the main text, we apply a rotation $U_x(\pi/2) = \exp(-i\sigma_x\pi/4)$ to the qubit initialized in $\ket{\downarrow}_z$ prior to the simulation (see Fig. \ref{figsequence} in the main text). This rotates the qubit to $\ket{\uparrow}_y$. At the end of the evolution, we apply another rotation $U_x(\pi/2)$ to perform the mapping $\ket{\uparrow}_y\leftrightarrow \ket{\uparrow}_z$ and $\ket{\downarrow}_y\leftrightarrow \ket{\downarrow}_z$. Therefore, to realize the electron transfer Hamiltonian in Eq. \eqref{eq_H} in the main text, the parameter mappings are $\Omega_y = \Delta E$, ${\Omega_x/2}=V_x$, $\eta\Omega = g$, and $\delta=-\omega$.

\subsection*{S2 System calibration} \label{SM_S2}

We independently calibrate the parameters of the laser tones used to realize both the unitary and the dissipative terms in Eq. \eqref{eq_master} in the main text. The spin-phonon coupling and phonon terms, $\frac{\eta\Omega}{2} \sigma_y (a^\dagger + a) - \delta a^\dagger a$, are calibrated by adjusting the Rabi frequency $\eta\Omega_n$ and detunings $\delta_n$ of the red and blue sideband Raman laser beatnotes from the tilt mode sideband resonances with $n=r,b$. 

We calibrate the spin-phonon coupling and the detuning by preparing the \Yb qubit in the $z$ basis and applying the Hamiltonian in Eq. \eqref{eq_Hord} in the main text with $\eta\Omega$ as the spin-phonon Rabi coupling strength. The hopping period corresponds to $2\pi/|\delta|$, which we use to estimate $\delta$. We then drive each tone on resonance to the tilt mode separately while setting the other tone off-resonant to estimate the effective $\eta\Omega$ for the experiments (see Figs. \ref{figA5}A-B).
In order to compensate for the ac-Stark shift due to the off-resonant excitation of the carrier transition by the red and blue sidebands, we use the following procedure: We first balance the Rabi coupling strengths of the red and blue sideband resonant drives ($\delta_n = 0$) to the tilt mode separately.  Then we turn on both tones simultaneously with the same detuning, $\delta_r =\delta_b = \delta$, from the tilt mode resonances. Using a Ramsey sequence, we adjust the ratio of the powers and a common frequency shift of the two laser tones to compensate for the undesired ac-Stark shift up to 0.25 kHz accuracy.
For the $\Omega_x$ and $\Omega_y$ Rabi frequencies in Eq. \eqref{eq_interactionHion}, we adjust the power of the two carrier transition tones independently (see Fig. \ref{figA5}C).


\begin{figure}
    \centering
    \includegraphics[width=\linewidth]{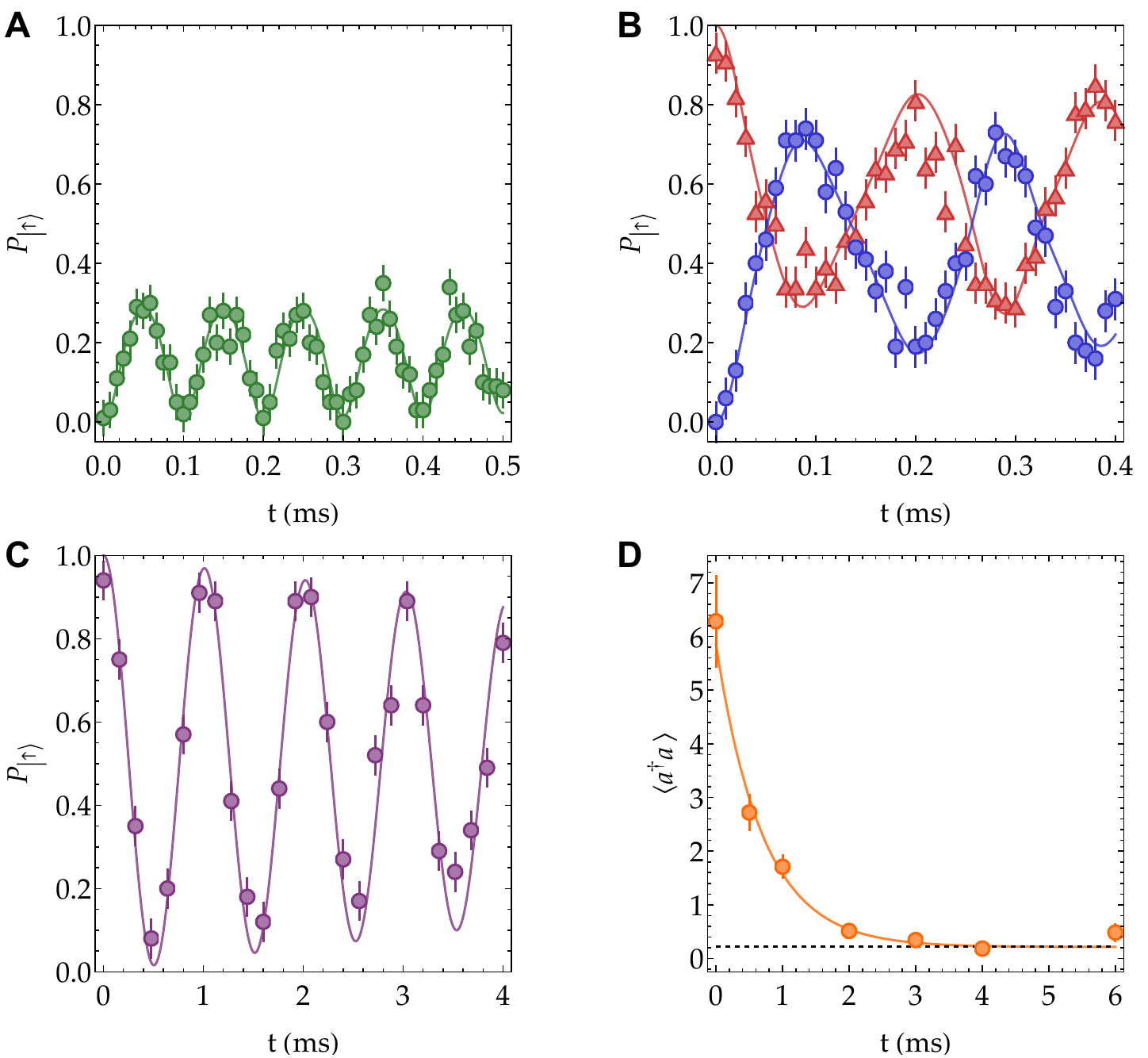}
    \caption{{\bf Hamiltonian and dissipation experimental calibration.}\\({\bf A}) Spin dynamics from the red and blue sideband Raman laser beatnotes with a common detuning from the tilt mode, $\delta/2\pi=-10~\,\rm kHz$, and equal Rabi coupling strengths, $\eta\Omega =0.55|\delta|$. The hopping period corresponds to $2\pi/|\delta|$.
    ({\bf B}) The same spin-phonon drives with one tone on resonance and another detuned from the tilt mode. Red triangles (experimental data) and solid curve (numerics) correspond to $\{\delta_r/2\pi,\delta_b/2\pi\} = \{0,-10\}$ kHz, and the blue circles correspond to $\{\delta_r/2\pi,\delta_b/2\pi\} = \{-10,0\}$ kHz. ({\bf C}) Spin dynamics undergoes a carrier drive along $x$ in the $\sigma_y$ basis. The Rabi coupling strength is set to $\Omega_x/2\pi = 0.99$ kHz. Together with another tone of the same frequency beatnote and phase difference of $\pi/2$, we generate the spin operation terms in the electron transfer Hamiltonian. The numerical results represented by solid curves in {(A)-(C)} include spin decoherence ($\gamma_z/2\pi = 10$ Hz) and motional dephasing ($\gamma_m/2\pi = 5$ Hz). ({\bf D}) The evolution of the average tilt mode phonon occupation number of the dual-species ion crystal via continuous resolved sideband cooling on \Ybc with 435 nm and 935 nm beams. The exponential constant determines the cooling rate. Here, $\gamma/2\pi = 0.23$ kHz, and the steady state average phonon occupation number is $\bar{n}=0.2$ 
    {(dashed horizontal line)}.
    }
    \label{figA5}
\end{figure}

The frequency of the 435 nm red sideband resonance of the \Ybc optical transition is found by using a scheme similar to Quantum logic spectroscopy (QLS) via spin-state measurements of \Yb. The pulse sequence consists of a series of three $\pi$-pulses on both ions, namely, BSB$\pi (355 {\rm nm}) \rightarrow$ RSB$\pi (435 {\rm nm}) \rightarrow$ BSB$\pi$(355 nm), while varying the frequency of the 435 nm RSB$\pi$ pulse. The 355 nm light is kept on and out of resonance with the \Yb qubit during 435 nm illumination to account for the differential stark shift on the 435 nm cooling transition. Another method to quantify the 435 nm cooling transition frequency is to replace the initial sideband cooling (SBC) pulses of the tilt mode with a finite amount of 435 nm continuous sideband cooling (CSBC) pulse. We then scan the RSB frequency of both the Zeeman $\Delta m_j=0$ transitions of \Ybc parking at $2\pi$ time of the tilt mode BSB evolution. By observing the contrast of the BSB population, we estimate the stark shifted frequencies of the 435 nm RSB pulse used for sympathetic cooling.

\begin{figure}[h]
    \centering
    \includegraphics[width=\linewidth]{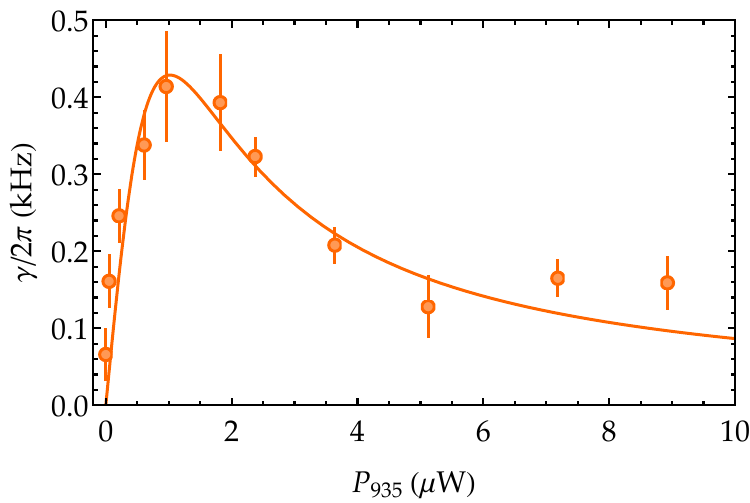}
    \caption{
    {\bf Cooling rate versus 935 nm power.} 
    \\An optimal 935 nm power is observed with the measured RSB Rabi frequency of about 3.4 kHz for each 435 nm tone. The solid curve is the theoretical results using the steady state solution of the master equation of a simplified three-level system ($\ket{g}$, $\ket{o}$, and $\ket{e}$) with $\gamma \approx \Gamma_{935}\rho_{ee}$, where $\Gamma_{935}$ is the scattering rate of $\ket{e}$ and $\rho_{ee}$ its the steady-state population. Here, we use the 935 nm detuning from $\ket{e}$, $\Delta_\text{935} = 2\pi$ $\times$ 6.4 MHz for the theory. A cooling rate baseline of $2\pi$ $\times$ 0.066 kHz is observed due to the residual cooling from the 935 nm tone of \Yb that is constantly on during the experiments.}
    \label{fig935}
\end{figure}

To estimate the cooling rate 
{$\gamma$ and the steady-state phonon $\bar{n}$} with 435 nm and 935 nm beams on \Ybc, we carry out the following procedure: we first perform Doppler cooling on the dual-species chain; then, we employ continuous sideband cooling on the tilt mode through \Ybc, followed by Raman sideband cooling on the com mode through \Yb; subsequently, we optically pump the spin state of 
\Yb to $\ket{\downarrow}_z$ and perform a phonon distribution measurement on tilt mode via a resonant BSB drive to estimate the average tilt mode phonon. By varying the cooling time and measuring the corresponding average phonon, we can obtain the cooling rate $\gamma$ and the average phonon $\bar{n}$ in Eq. \eqref{eq_master} in the main text with an exponential fit, as shown in Fig. \ref{figA5}D. The cooling rate can be adjusted by changing the 935 nm repumper power as it is non-monotonically dependent on the 935 nm power exhibiting an optimum, as shown in Fig. \ref{fig935}. 

\subsection*{S3 Numerical simulations} \label{SM_S3}

We simulate Eq. \eqref{eq_master} in the main text using a Python package based on QUTIP \cite{JOHANSSON2013}, which allows us to include experimental imperfections that induce different types of dephasing in our system.  As the experiment is performed in the rotated basis ($z\leftrightarrow y$), fluctuations in the laser intensity and detuning cause effective spin decoherence, while trap frequency fluctuations and the heating rate of tilt motional mode cause motional dephasing. Therefore, when comparing the numerics with the experimental results, we introduce two additional dissipative processes, which modify Eq. \eqref{eq_master} in the main text to
\begin{equation}
    \frac{\partial\rho}{\partial t}=-i[H_{\rm s},\rho] + \gamma (\bar{n}+1)\mathcal{L}_{a}[\rho] + \gamma \bar{n} \mathcal{L}_{a^\dagger}[\rho] + \sum_{k=z,m}\gamma_k\mathcal{L}_{c_k}[\rho],
    \label{eq_master_mod}
    \tag{S.9}
\end{equation}
where the jump operator $c_z=\sigma_y$ and its corresponding rate $\gamma_z$ account for spin decoherence while the jump operator $c_m=a a^\dagger + a^\dagger a$ and its corresponding rate $\gamma_m$ account for motional dephasing \cite{Fluhmann2019encoding}. We determine these dephasing rates by comparing numerical results to experimental data, finding that $\gamma_z/2\pi\sim (0-10)$ Hz, and $\gamma_m/2\pi \sim 5$ Hz.

\subsection*{S4 Electron transfer transient dynamics} \label{SM_S4}

As discussed in the main text, the spin evolutions in the nonadiabatic and strongly adiabatic regimes exhibit different behaviors. In Fig. \ref{figEvolution}, we report the time traces of the donor population evolution for $\Delta E=\omega/2, \omega, 2\omega$ in both regimes. As shown in Figs. \ref{figEvolution}A-B, the dynamics of the donor population in the nonadiabatic regime can be modeled with an exponential decay. On the other hand, the dynamics in the strongly adiabatic regime feature complex oscillatory decays that cannot be fitted with an exponential decay as those in the nonadiabatic regime (see Figs. \ref{figEvolution}C-D). In the strongly adiabatic regime, when $\Delta E=\omega$, the oscillations are evident because the effective vibronic coupling is dominated by the $V_x\sqrt{\text{FC}_{0_-,1_+}}$ coupling term, which has the largest Franck-Condon factor. Meanwhile, in the $\Delta E=\omega/2$ case, the oscillations are less pronounced because of the non-integer value of $\Delta E/\omega$ that suppresses the transfer, analogous to off-resonant coupling in a two-level system (see Fig. \ref{figEvolution}C). Finally, when $\Delta E=2\omega$, the relevant Franck-Condon factor $\rm FC_{0_-,2_+}$ is smaller than the $\Delta E=\omega$ case, which decreases the effective vibronic coupling, resulting in both smaller frequency and amplitude of the oscillations.

\begin{figure}[h!]
    \centering
    \includegraphics[width=\linewidth]{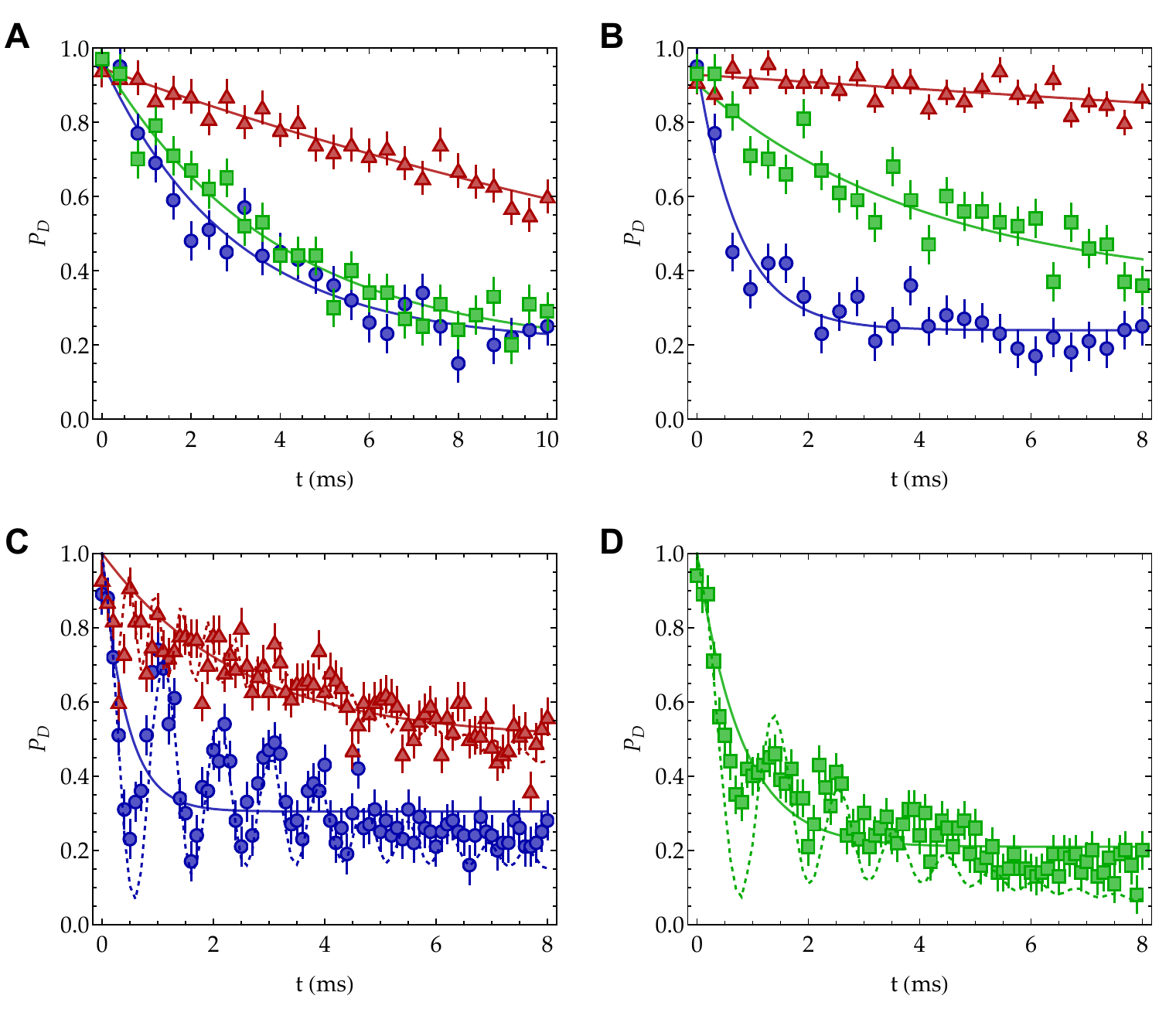}
    \caption{
    {\bf Donor population dynamics.} \\({\bf A} to {\bf D}) Time-evolution traces of the donor population in the (A and B) nonadiabatic (from Figs. 3B and 3E) and {(C and D)} adiabatic regime (from Fig. 4B). Red triangles, blue circles, and green squares are experimental data for $\Delta E = \omega/2$, $\Delta E = \omega$, and $\Delta E = 2\omega$, respectively. The solid curves are their corresponding exponential fits. The dynamics in the nonadiabatic regime is well described by an exponential decay. Conversely, in the strongly adiabatic regime, the donor population exhibits initial oscillations between the donor and acceptor states, and the exponential fit overestimates both the transfer rate and the final donor population. In (C and D), we also include the numerical results from Eq. \eqref{eq_master} in the main text to highlight the oscillatory behavior of the strongly adiabatic regime.}
    \label{figEvolution}
\end{figure}

\subsection*{S5 Phonon steady state characterization} \label{SM_S5}

In addition to the spin degrees of freedom, a trapped-ion platform has the inherent capability of measuring motional observables, allowing access to spin-phonon correlations as well as to models in which the motional degrees of freedom govern the dynamics. For example, the observations of the destructive interference of the wave packet due to the geometric phase in the quantum simulation of conical intersections are performed through the reconstruction of the ion's wave function spatial distribution \cite{Whitlow2023,Valahu2023}.

Although the motional degree of freedom only plays a surrogate role in the dynamics of the electron transfer model studied here, we note that, from a quantum optics perspective, our system simulates a variant of the Rabi model \cite{Lv2018} with tunable dissipation, ranging from weak to ultrastrong coupling regimes. In this regard, the Rabi model with dissipation can be investigated by measuring both motional and spin observables. In this work, we study the electron transfer models in the $g\gg\gamma$ regime, where the dynamics of the spin degree of freedom is inherently non-Markovian. Therefore, the spin and the oscillator can have finite correlations in the steady state \cite{huelga2012non}.
Specifically, in the nonadiabatic regime at $\Delta E = 0$, the steady state of the ET system is a classical ensemble of donor and acceptor vibronic states with equal weights. 
Meanwhile, the steady state in the strongly adiabatic regime at $\Delta E=0$ is a quantum superposition of donor and acceptor vibronic states with equal probabilities.
As $\Delta E$ increases, the ensemble weight or quantum probability of the acceptor vibronic state increases towards $1$ in both cases. 
Such a change in the steady-state structure can be captured by a decrease in the spin-phonon correlation, which in turn affects also the steady-state phonon population.
Here, we measure the average phonon population in the steady state of the evolution under Eq. \eqref{eq_master} in the main text and indirectly observe spin-phonon correlations depending on the donor-acceptor energy separation $\Delta E$.

First, we discuss the steady state of the Lindbladian master equation, focusing in particular on the properties of the phonon population. From Eq. \eqref{eq_master} in the main text, the expectation of an observable $O$ satisfies
\begin{align}
    \partial_t\langle O\rangle = & \, {\rm Tr} \left[-iO\left[H,\rho\right]+O\gamma\left(\bar{n}+1\right)\mathcal{D}_a\left(\rho\right)\right.\nonumber\\&\left.+O\gamma\bar{n}\mathcal{D}_{a^\dag}\left(\rho\right)\right]\nonumber\\
    =&-i\langle\left[O,H\right]\rangle+\gamma\left(\bar{n}+1\right)\langle a^\dag Oa-\frac{1}{2}\left\{O,a^\dag a\right\}\rangle\nonumber\\
    &+\gamma\bar{n}\langle aOa^\dag-\frac{1}{2}\left\{O,aa^\dag\right\}\rangle.
    \tag{S.10}
\end{align}
Using the bosonic commutation relation, the number operator $n=a^\dag a$ satisfies
\begin{equation}
\partial_t\langle n\rangle=-i\frac{g}{2}\langle\sigma_z\left(a^\dag-a\right)\rangle+\gamma\left(\bar{n}-\langle n\rangle\right).
\label{n_eq}
\tag{S.11}
\end{equation}
The creation/annihilation operators satisfy
\begin{align}
    \partial_t\langle a^\dag\rangle&=i\frac{g}{2}\langle\sigma_z\rangle+\left(i\omega-\gamma/2\right)\langle a^\dag\rangle,
    \nonumber\\
    \partial_t\langle a\rangle&=-i\frac{g}{2}\langle\sigma_z\rangle-\left(i\omega+\gamma/2\right)\langle a\rangle.
    \label{a_eq}
    \tag{S.12}
\end{align}
To obtain steady-state solutions, we set LHS of Eq. \eqref{n_eq}, \eqref{a_eq} equal to zero leading to
\begin{equation}
    n_{\rm ss}=\bar{n}-\frac{i}{2}\frac{g}{\gamma}\langle\sigma_z\left(a^\dag-a\right)\rangle_{\rm ss}, \label{n_ss}
    \tag{S.13}
\end{equation}
\begin{equation}
    \langle a^\dag\rangle_{\rm ss}=-\frac{ig}{2i\omega-\gamma}\langle\sigma_z\rangle_{\rm ss},\;\; \langle a\rangle_{\rm ss}=-\frac{ig}{2i\omega+\gamma}\langle\sigma_z\rangle_{\rm ss}\label{a_ss},
    \tag{S.14}
\end{equation}
which immediately gives an expression of steady-state reaction coordinate $y_{\rm ss}$ in terms of the steady-state donor population $P_D^{\rm ss}$ as follows
\begin{equation}
    y_{\rm ss}=\frac{y_0}{2}\left(a+a^\dag\right)=-\frac{2\omega g}{4\omega^2+\gamma^2}(2P_D^{\rm ss}-1)y_0.
    \tag{S.15}
\end{equation}
To quantify spin-phonon correlation, we can compare the exact steady-state $n_{\rm ss}$ in Eq. \eqref{n_ss} with the one calculated assuming that spin and phonon are in an uncorrelated state, namely, $\langle\sigma_z\left(a^\dag-a\right)\rangle_{\rm ss}=\langle\sigma_z\rangle_{\rm ss}\langle a^\dag-a\rangle_{\rm ss}$, which leads to
\begin{equation}
n_{\rm ss}^{\rm un}=\bar{n}+\frac{g^2}{4\omega+\gamma^2}\left(2P_D^{\rm ss}-1\right)^2.
\label{n_mf}
\tag{S.16}
\end{equation}

\begin{figure}[t!]
    \centering
    \includegraphics[width=\linewidth]{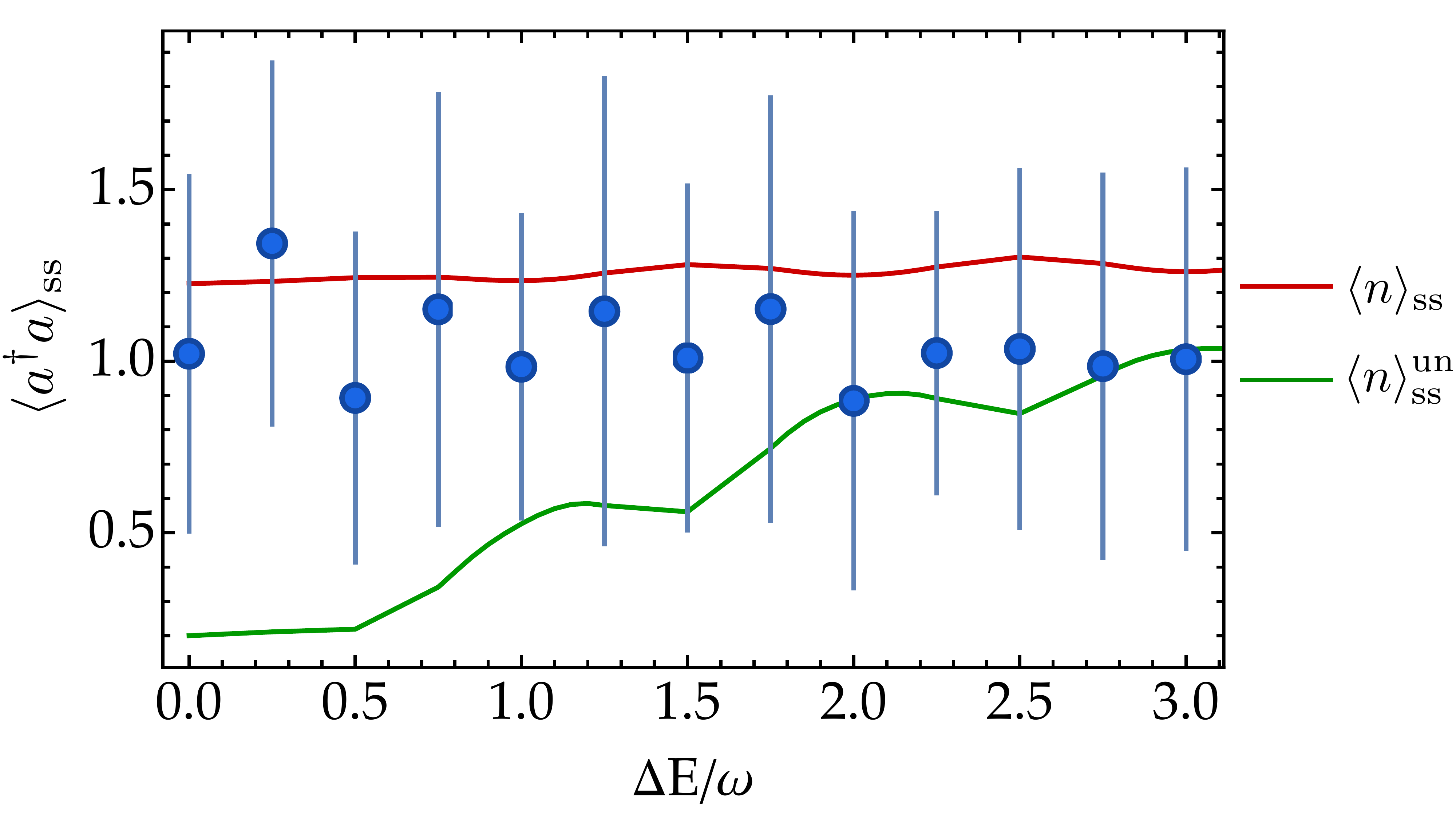}
    \caption{
    {\bf Characterization of steady-state phonon population.} \\Average phonon population in the steady state (blue circles) as a function of $\Delta E$ using $(V_x,g,\gamma)=(0.19,1.91,0.038)\omega$, $\delta/2\pi=-4~\rm kHz$, and $\bar{n}=0.2$. The average phonon population is extracted by fitting the first six phonon states. The error bars are the standard deviations from the mean. The dark red solid line is the exact prediction given by Eq. \eqref{n_ss} while the dark green solid line is the prediction given by Eq. \eqref{n_mf}. At low $\Delta E$, there is spin-phonon correlation in the steady state, which decreases monotonically as $\Delta E$ is increased. Here, we also consider a motional dephasing of $\gamma_m = 0.0013\omega$.}
    \label{fig5}
\end{figure}

In Fig. \ref{fig5}, we report measurements of the steady-state $\langle a^\dagger a \rangle_{\rm ss}$ as a function of $\Delta E$ 
{in the case of ultrastrong coupling ($g/\omega=1.91$)}. To measure the steady-state phonon population, the qubit is reset via optical pumping, followed by the application of a resonant BSB Hamiltonian $H^\text{BSB}=i(\eta\Omega/2)(a\sigma^- - a^\dagger\sigma^+)$ to the system after the evolution has reached its steady state. The resulting spin evolution is then fitted to extract the average phonon population $\langle n\rangle_{\rm ss}=\langle a^\dagger a \rangle_{\rm ss}$ in the steady state. The measured values are in agreement with the exact solution $\langle n \rangle_{\rm ss}$ in Eq. \eqref{n_ss}, confirming the presence of spin-phonon correlations in the system's steady state.

Furthermore, to get an intuitive understanding of the effect of the dissipation due to the Lindbladian, we shall assume the spin is either in $\ket{\uparrow}$ or $\ket{\downarrow}$ such that the Hamiltonian in Eq. \eqref{eq_H} in the main text can be reduced to
\begin{equation}
    H_p = \omega a^\dag a \pm \frac{g}{2}(a+a^\dag).
    \tag{S.17}
\end{equation}
The corresponding master equation becomes 
\begin{equation}
    \partial_t \rho=-i[H_p,\rho] + \gamma (\bar{n}+1)\mathcal{D}_{a}[\rho] + \gamma \bar{n} \mathcal{D}_{a^\dagger}[\rho].
    \label{ms_p}
    \tag{S.18}
\end{equation}
We can define displaced bosonic creation/annihilation operators $b\equiv a+\alpha, b^\dagger\equiv a^\dagger+\alpha^*$ with $\alpha$ being a complex constant to be determined. It can be shown that if we set
\begin{equation}
    \alpha = \pm \frac{2g\omega}{4\omega^2+\gamma^2}\pm\frac{ig\gamma}{4\omega^2+\gamma^2},
    \tag{S.19}
\end{equation}
Eq. \eqref{ms_p} then becomes
\begin{equation}
    \partial_t \rho=-i[\omega b^\dagger b,\rho] + \gamma (\bar{n}+1)\mathcal{D}_{b}[\rho] + \gamma \bar{n} \mathcal{D}_{b^\dagger}[\rho],
    \tag{S.20}
\end{equation}
which takes the form of a simple damped oscillator with steady state being a thermal vibrational state characterized by $\bar{n}$ \cite{Petruccione2007}. 
Undoing the displacement gives us the steady-state solution of Eq.\eqref{ms_p}:
\begin{equation}
    \rho_{ss} = D(-\alpha) \rho_{th} D(\alpha), \;  \rho_{th}=\frac{e^{-\beta\omega a^\dag a}}{1-e^{-\beta\omega}},
    \tag{S.21}
\end{equation}
where $D$ is the displacement operator and $1/\beta = \omega/\log(1 + 1/\bar{n})$. In the experiment, $\omega\gg\gamma$ such that $\alpha\rightarrow \pm g/2\omega$. Depending on the spin, the system is effectively pumped to the ground state of the left/right displaced harmonic well, as shown in Fig. \ref{fig1}B.


\subsection*{S6 Ohmic bath and Lindbladian formalism}\label{app_Lindblad}
In this section, we show that the derivation of a Lindbladian master equation for the system in Eq. \eqref{eq_H} in the main text in contact with an Ohmic bath is equivalent to the dissipative spin-boson model realized in this experiment under certain conditions. More formally, we will establish the equivalence between Eq. \eqref{eq_master} in the main text and the spin-boson Hamiltonian $H_{ET}$ in Eq. (1.3) of \cite{Garg1985}, namely,
\begin{equation}
    H_{ET} = H_{\rm s}+H_{\rm sb}+H_{\rm b}.
    \tag{S.22}
\end{equation}
The system Hamiltonian $H_{\rm s}$ is given by Eq. \eqref{eq_H} in the main text. The bath Hamiltonian $H_{\rm b}=\sum_{n}{\omega_n\Gamma_n^\dag\Gamma_n}$ is described by a collection of infinite harmonic oscillators with $\Gamma_n(\Gamma^\dagger_n)$ being the annihilation(creation) operator of the $n$-th bosonic mode. The reaction coordinate of the system is linearly coupled to the position coordinate of the bath via
\begin{equation}
    H_{\rm sb}=S\otimes B,\:S\equiv a+a^\dag, \:\: B\equiv K+K^\dag,
    \label{Hsb}
    \tag{S.23}
\end{equation}
 with $K^\dag\equiv\ \sum_{n}{c_n\Gamma_n^\dag}$ being a linear combination of bath operators and $c_n$ being the coupling coefficients of the $n$-th mode. The coupling coefficients $c_n$ and the bath frequencies determine the bath spectral density function $J\left(\omega\right)=\sum_{n}{c_n^2\delta\left(\omega-\omega_n\right)}$. We take $J(\omega)$ to be Ohmic \cite{Garg1985}:
\begin{equation}
    J(\omega) = \eta\omega \exp(-\omega/\omega_c), \,\, \omega_c\rightarrow\infty,
    \tag{S.24}
\end{equation}
which corresponds to a classical damped oscillator with $\eta$ being the linear damping coefficient. Note that in this section, we use $\omega$ as the frequency variable for the spectral density function $J$ and $\omega_0$ as the bosonic mode frequency in Eq. \eqref{eq_H} in the main text. 

To obtain the reduced dynamics of the system density matrix $\rho(t)$, we shall first change into the interaction picture of $H_0=H_{\rm s}+H_{\rm b}$. Let us denote a generic operator $O$ in the interaction picture of $H_0$ as
\begin{equation}
    \widetilde{O}(t)=U^\dag(t)OU(t),\:\:
     U(t)=\exp{\left(-i H_{0}t\right)},
     \label{int_pic}
     \tag{S.25}
\end{equation} 
the master equation in the interaction picture is then given by
\begin{equation}
\partial_t\widetilde{\chi}\left(t\right)=-i[{\widetilde{H}}_{\rm sb}(t),\widetilde{\chi}\left(t\right)],
\tag{S.26}
\end{equation}
where $\chi$ is the system-bath density matrix.  Explicitly integrating this equation and inserting the expression for $\widetilde{\chi}$ back leads to
\begin{equation}
\partial_t\widetilde{\chi}(t)=-i[{\widetilde{H}}_{\rm sb}\left(t\right),\chi\left(0\right)]-\int_{0}^{t}{dt^\prime[{\widetilde{H}}_{\rm sb}(t),[{\widetilde{H}}_{\rm sb}(t^\prime),\chi\left(t^\prime\right)]]}. \label{ms_int_0}
\tag{S.27}
\end{equation}

Assuming the system-bath coupling is sufficiently weak, and the bath is kept at thermal equilibrium, $\chi(t)$ becomes separable (Born approximation)
\begin{equation}
    \chi(t)=\rho(t)\otimes R_0, \:\:\: R_0 = e^{-\beta H_{\rm b}}/{\rm Tr}( e^{-\beta H_{\rm b}}),
    \label{BornA}
    \tag{S.28}
\end{equation}
where $\beta=1/k_B T$. Taking partial trace with respect to the bath degrees of freedom on Eq. \eqref{ms_int_0} gives an equation for $\widetilde{\rho}$
\begin{align}
\partial_t\widetilde{\rho}\left(t\right)=&-\int_{0}^{t}{dt^\prime}
C\left(\tau\right) \left[\widetilde{S}\left(t\right)\widetilde{S}\left(t^\prime\right)\widetilde{\rho}\left(t^\prime\right)-\widetilde{S}\left(t^\prime\right)\widetilde{\rho}\left(t^\prime\right)\widetilde{S}\left(t\right)\right] \nonumber \\
&+C\left(-\tau\right)\left[\widetilde{\rho}\left(t^\prime\right)\widetilde{S}\left(t^\prime\right)\widetilde{S}\left(t\right)-\widetilde{S}\left(t\right)\widetilde{\rho}\left(t^\prime\right)\widetilde{S}\left(t^\prime\right)\right],
\label{ms_ex}
\tag{S.29}
\end{align}
where $\tau\equiv t-t'$, and $C(\tau)$ is the temporal correlation function of the bath, namely,
\begin{align}
C\left(\tau\right)&={\rm Tr}\left[\widetilde{B}\left(\tau\right)BR_0\right] \nonumber\\
&=\int_{0}^{\infty}d\omega J\left(\omega\right)\left[\coth{\left(\beta\omega/2\right)}\cos{(\omega \tau)}-i\sin{(\omega \tau)}\right].\nonumber
\label{Ctau}
\tag{S.30}
\end{align}
$S,\widetilde{S},B,\widetilde{B}$ follow the definitions in Eq. \eqref{Hsb},\eqref{int_pic}.\\
By introducing a displaced bosonic operator $b=a+a_0$ with the spin dependent constant $a_0=\frac{g}{2\omega_0}\sigma_z$, we can evaluate the interaction frame system operator $\tilde{S}(t)$
\begin{equation}
\widetilde{S}\left(t\right)=\left(a^\dag e^{i\omega_0t}+ae^{-i\omega_0t}\right)+2a_0\left(\cos{\omega_0t}-1\right).
\label{S_int}
\tag{S.31}
\end{equation}
Assuming the bath correlation function is strongly peaked around $\tau=0$ with a correlation time $\tau_r$ much smaller than the typical time scale of the system’s dynamics, $C(\tau)$ can be approximated as $\delta(\tau)$, which yields the replacement $\rho\left(t^\prime\right)\rightarrow\rho\left(t\right)$. Another important effect of this approximation is the extension of the integration limit from $t$ to $\infty$ of the integral in Eq. \eqref{ms_ex} (Markovian approximation) \cite{carmichael2013statistical}.  Eq. \eqref{ms_ex} then results in the Schr\"{o}dinger picture master equation
\begin{align}
    \partial_t\rho\left(t\right)=&-i\left[H_{\rm s},\rho\left(t\right)\right]\nonumber\\ 
     &-\int_{0}^{\infty}d\tau C\left(\tau\right)\left[S\widetilde{S}\left(-\tau\right)\rho\left(t\right)-\widetilde{S}\left(-\tau\right)\rho\left(t\right)S\right]\nonumber \\
      &+ \int_{0}^{\infty}d\tau C\left(-\tau\right)\left[\rho\left(t\right)\widetilde{S}\left(-\tau\right)S-S\rho\left(t\right)\widetilde{S}\left(-\tau\right)\right].
       \label{ms_s0}
       \tag{S.32}
\end{align}
Inserting Eq. \eqref{Ctau} and Eq. \eqref{S_int}, this equation can be written in a compact form as

\begin{align}
      \partial_t\rho\left(t\right)=&-i\left[H_{\rm s},\rho\left(t\right)\right]\nonumber\\ &- \left[S(\Lambda+C_0)\rho\left(t\right)-(\Lambda+C_0)\rho\left(t\right)S
     +{\rm h.c.}\right],
     \label{ms_s}
     \tag{S.33}
\end{align}
where $\Lambda=\mathcal{L}_+a+\mathcal{L}a^\dag$ and $\mathcal{L}_+,\mathcal{L}$ take form of Laplace transform
  \begin{align}
      \mathcal{L}_+&=\int_{0}^{\infty}{d\omega J\left(\omega\right)\left[1+\bar{n}\left(\omega\right)\right]\int_{0}^{\infty}{d\tau e^{-i\left(\omega-\omega_0\right)\tau}}},\nonumber\\
      \mathcal{L}&=\int_{0}^{\infty}{d\omega J\left(\omega\right)\bar{n}\left(\omega\right)\int_{0}^{\infty}{d\tau e^{i\left(\omega-\omega_0\right)\tau}}}\label{Lap},
      \tag{S.34}
\end{align}
where $\bar{n}(\omega)$ is the thermal phonon population of a mode of frequency $\omega$, and $C_0$ is a constant due to the scalar part of Eq. \eqref{S_int},
\begin{equation}
    C_0 = 2a_0\int_{0}^{\infty}d\tau C\left(\tau\right)\left(\cos{\omega_0\tau}-1\right).\label{c_cons}
    \tag{S.35}
\end{equation}
It shall be noted that in the evaluation of $\Lambda$, we have applied the secular approximation, neglecting the contribution from highly oscillatory terms involving $e^{\pm i(\omega+\omega_0)t}$ \cite{brandes2004chapter}. 

Evaluating the integrals in Eq. \eqref{Lap}, \eqref{c_cons} leads to the reduced master equation of the system
\begin{align}
\partial_t\rho=&-i\left[H_{\rm s}+H_c,\rho\right]-i\left[H_n\rho-\rho H_n^\dag\right]\nonumber\\
    &+\gamma\left({\bar{n}}_0+1\right)\left(\mathcal{D}_a(\rho)+\mathcal{D}_a'(\rho)\right) \nonumber\\
    &+\gamma{\bar{n}}_0\left(\mathcal{D}_{a^\dag}(\rho)+\mathcal{D}_{a^\dag}^\prime(\rho)\right)\nonumber\\
    &+i\Delta_d\mathcal{D}_a^{Im}(\rho).
    \label{eq_master_D}
    \tag{S.36}
\end{align}
Let us break down the different terms of the master equation \eqref{eq_master_D}: $\mathcal{D}_c(\rho)$ is the Lindbladian super-operator defined in Eq. \eqref{eq_Lind} in the main text with coefficients,
\begin{equation}
    \gamma=2\pi\eta\omega_0,\:\:\bar{n}_0=\bar{n}(\omega_0).
    \tag{S.37}
\end{equation}
$\mathcal{D}_c^\prime(\rho)$ is defined as
\begin{equation}
\mathcal{D}_c^\prime(\rho)\equiv\frac{1}{2}\left(c^\dag\rho c^\dag+c\rho c-c^\dag c^\dag\rho-\rho cc\right),
\tag{S.38}
\end{equation}
$\mathcal{D}_a^{Im}\left(\rho\right)$ represents a superoperator with imaginary coefficients
\begin{equation}
\mathcal{D}_c^{Im}\left(\rho\right)\equiv\left(c\rho c-c^\dag\rho c^\dag\right),
\tag{S.39}
\end{equation}
$H_n$ represents the following non-hermitian Hamiltonian
\begin{equation}
    H_n=\Delta_d a a, \, \Delta_d = P\left[\int_{0}^{\infty}{d\omega\frac{(2\bar{n}\left(\omega\right)+1)J\left(\omega\right)}{\omega_0-\omega}}\right],
    \tag{S.40}
\end{equation}
where $P$ stands for Principal Value, and $H_c$ is a correction to the system Hamiltonian
\begin{align}
    H_c&=\widetilde{\omega}a^\dag a+\frac{\widetilde{g}}{2}\sigma_z\left(a+a^\dag\right),\nonumber\\
    \widetilde{\omega}&=P\left[\int_{0}^{\infty}{d\omega\frac{J\left(\omega\right)}{\omega_0-\omega}}\right],\nonumber\\
    \widetilde{g}&=4gP\left[\int_{0}^{\infty}{d\omega\frac{J\left(\omega\right)}{\omega_0^2-\omega^2}}\ \right].\nonumber
    \tag{S.41}
\end{align}
When the frequency of the system's oscillator $\omega_0$ is much larger than the decay rate $\gamma$, under rotating-wave approximation, we can effectively neglect the terms involving $aa,a^\dag a^\dag$ that do not conserve the energy \cite{carmichael2013statistical}. These include the superoperators $\mathcal{D}_a^\prime(\rho),\mathcal{D}_{a^\dag}^\prime(\rho),\mathcal{D}_a^{Im}\left(\rho\right)$, and the non-hermitian Hamiltonian terms $ H_n$. The rotating-wave approximation is consistent with the Born approximation, which assumes that the system-bath coupling is sufficiently small so that the system and bath can be described by a separable state. After the above approximations, \eqref{eq_master_D} takes form of a standard Lindbladian master equation \cite{mohseni2014quantum, nielsen2001quantum}, and we obtain Eq. \eqref{eq_master} in the main text, with renormalized oscillator frequency $\omega'=\omega_0+\widetilde{\omega}$ and displacement $g'=g+\widetilde{g}$.

We can check the validity of the Markovian approximation by comparing the time scale of the system dynamics $\tau_{\rm s}~\sim~1/\gamma$ and the width of the position correlation function estimated by the bath correlation time $\tau_r\sim\beta$ \cite{carmichael2013statistical}. For the approximation to be valid, we therefore require $\tau_r\ll\tau_{\rm s}$. Hence, for Eq. \eqref{eq_master} in the main text to be a good description of the ET model in the weak decay regime, we require the following conditions, which are also derived in \cite{Lemmer2018}:
\begin{align}
     \gamma&\ll1/\beta, \tag{Markovian}\\
     \gamma &\ll \omega_0 \tag{RWA, Born}.
\end{align}

\end{document}